\documentclass[twocolumn,showpacs,preprintnumbers,superscriptaddress,amsmath,amssymb,nofootinbib]{revtex4}

\usepackage{graphicx}
\usepackage{dcolumn}
\usepackage{bm}
\usepackage{amsmath, amssymb}

\usepackage[normalem]{ulem}  
\usepackage[dvipdfmx]{color} 

\renewcommand\sout{\bgroup \color{red} \ULdepth=-.5ex \ULset}

\newcommand{\Slash}[1]{\ooalign{\hfil/\hfil\crcr$#1$}}
\newcommand{\Psfig}[2]{\includegraphics[width=#1]{#2}}
\newcommand{\PsfigII}[2]{\includegraphics[scale=#1]{#2}}

\newcommand{\one}{\mbox{1}\hspace{-0.25em}\mbox{l}}

\def\Kaellen{K\"{a}llen }
\def\Schr{Schr\"{o}dinger }

\def\mev{\text{ MeV}}
\def\gev{\text{ GeV}}

\def\prt{\partial}

\begin{document}

\preprint{}

\title{Compositeness of baryonic resonances: \\*
  Applications to the $\bm{\Delta (1232)}$, $\bm{N (1535)}$, and
  $\bm{N (1650)}$ resonances}

\author{Takayasu Sekihara} 
\email{sekihara@rcnp.osaka-u.ac.jp}
\affiliation{Research Center for Nuclear Physics
  (RCNP), Osaka University, Ibaraki, Osaka, 567-0047, Japan}

\author{Takashi Arai} 
\affiliation{Theory Center, IPNS, High Energy Accelerator Research
  Organization (KEK), \\ 1-1 Oho, Tsukuba, Ibaraki 305-0801, Japan}

\author{Junko Yamagata-Sekihara} 
\affiliation{National Institute of Technology, Oshima College, Oshima,
  Yamaguchi, 742-2193, Japan}

\author{Shigehiro Yasui}
\affiliation{Department of Physics, Tokyo Institute of Technology,
  Tokyo 152-8551, Japan}

\date{\today}

\begin{abstract}

  We present a formulation of the compositeness for baryonic
  resonances in order to discuss the meson--baryon molecular structure
  inside the resonances.  For this purpose, we derive a relation
  between the residue of the scattering amplitude at the resonance
  pole position and the two-body wave function of the resonance in a
  sophisticated way, and we define the compositeness as the norm of
  the two-body wave functions.
  As applications, we investigate the compositeness of the $\Delta
  (1232)$, $N (1535)$, and $N (1650)$ resonances from precise $\pi N$
  scattering amplitudes in a unitarized chiral framework with the
  interaction up to the next-to-leading order in chiral perturbation
  theory.  The $\pi N$ compositeness for the $\Delta (1232)$ resonance
  is evaluated in the $\pi N$ single-channel scattering, and we find
  that the $\pi N$ component inside $\Delta (1232)$ in the present
  framework is nonnegligible, which supports the previous work.  On
  the other hand, the compositeness for the $N (1535)$ and $N (1650)$
  resonances is evaluated in a coupled-channels approach, resulting
  that the $\pi N$, $\eta N$, $K \Lambda$ and $K \Sigma$ components
  are negligible for these resonances.

\end{abstract}

\maketitle

\section{Introduction}

Investigating the internal structure of hadrons is one of the most
important topics in hadron physics~\cite{Olive:1900zz}, highly
motivated by our expectation that there can exist exotic hadrons,
which are not composed of a three-quark ($q q q$) system for baryons
nor of a quark--antiquark ($q \bar{q}$) one for mesons.  Namely, while
traditional quark models have succeeded in describing baryons and
mesons with $q q q$ and $q \bar{q}$, respectively, we may consider
some exotic configurations for hadron structures, {\it e.g.},
tetraquarks and pentaquarks, as long as they are color singlet states.
Indeed, there are several candidates of exotic hadrons, which cannot
be classified into the states predicted by the quark models.  For
instance, $\Lambda (1405)$ has been considered as an exotic hadron
rather than a compact $u d s$ state because of its anomalously light
mass; since the $\bar{K} N ( I = 0 )$ interaction is strongly
attractive, $\Lambda (1405)$ may be a $\bar{K} N$ molecular
state~\cite{Dalitz:1960du}.  Until now, great efforts have been
continuously made in both experimental and theoretical sides in order
to clarify the structure of exotic hadron candidates and to discover
genuine exotic hadrons.  In this context, it is encouraging that there
have been experimental signals of exotic hadrons in the heavy quark
sector: charged charmonium-like states discovered by
Belle~\cite{Belle:2011aa} and charmonium pentaquarks by
LHCb~\cite{Aaij:2015tga}.  Moreover, it is interesting that an
evidence of $\Lambda (1405)$ as a $\bar{K} N$ molecular state has come
from lattice QCD simulations~\cite{Hall:2014uca}.

Among exotic configurations of hadrons, hadronic molecular states are
of special interest, since they are composed of two or more asymptotic
states of QCD, i.e., color singlet states, and hence one can define
the structure of these hadrons in hadronic degrees of freedom without
complicated treatment of QCD.  Actually, because hadrons are color
singlet states, their masses and interactions between them do not
depend on the renormalization scheme of QCD, in contrast to the
quark--gluon dynamics.  This viewpoint of the study on composites of
asymptotic states originates in the old work on the field
renormalization constant intensively discussed in the
1960s~\cite{Salam:1962ap, Weinberg:1962hj, Ezawa:1963zz, Lurie:1963}.
One of the most prominent results in this approach is that the
deuteron is dominated by the loosely bound proton--neutron
component~\cite{Weinberg:1965zz}.  Then, the study of the structure of
hadrons from the field renormalization constant and from the so-called
compositeness has been developed in, {\it e.g.},
Refs.~\cite{Baru:2003qq, Hyodo:2011qc, Aceti:2012dd, Xiao:2012vv,
  Hyodo:2013iga, Hyodo:2013nka, Hanhart:2014ssa, Hyodo:2014bda,
  Aceti:2014ala, Aceti:2014wka, Nagahiro:2014mba, Sekihara:2014kya,
  Garcia-Recio:2015jsa, Guo:2015daa, Kamiya:2015aea}.  In particular,
the compositeness is explicitly defined as contributions from two-body
wave functions to the normalization of the total wave function for the
resonance~\cite{Aceti:2012dd, Sekihara:2014kya, Gamermann:2009uq,
  YamagataSekihara:2010pj}, and can be extracted from the scattering
amplitude for two asymptotic states.  Since the total wave function is
normalized to be unity, we can discuss the composite fraction of
hadrons by comparing the compositeness with unity.  Here we should
note that, in general, the compositeness as well as the wave function
is not observable and hence a model dependent quantity.  However, for
states lying near the two-body threshold, we can express the
compositeness with observables such as the scattering length and
effective range, as studied in Refs.~\cite{Weinberg:1965zz,
  Baru:2003qq, Hyodo:2013iga, Hanhart:2014ssa, Hyodo:2014bda,
  Kamiya:2015aea}.  Besides, in a certain model the compositeness has
been utilized to study the internal structure of hadronic resonances
from experimental observable as well, such as the $\bar{K} N$
component inside $\Lambda (1405)$~\cite{Sekihara:2013sma} and the $K
\bar{K}$ components inside the scalar mesons $f_{0} (980)$ and $a_{0}
(980)$~\cite{Sekihara:2014qxa}.

Since the compositeness can be extracted from the scattering
amplitude, it is a good subject to apply the compositeness to the
nucleon resonances, which we abbreviate as $N^{\ast}$, and to discuss
the meson--baryon compositeness for the $N^{\ast}$ resonances.  This
is because, at present, precise $\pi N$ scattering amplitudes are
available by many research groups, {\it e.g.},
ANL--Osaka~\cite{Kamano:2013iva}, J\"{u}lich~\cite{Ronchen:2012eg},
and Dubna--Mainz--Taipei~\cite{Chen:2007cy} in the so-called dynamical
approaches, and Bonn--Gatchina~\cite{Anisovich:2011fc} and
GWU~\cite{Workman:2012hx} in the on-shell $K$-matrix approaches.  In
principle we can extract the $\pi N$ and other meson--baryon
compositeness from the precise $\pi N$ scattering amplitudes via
properties of the resonance poles.

In this paper we focus on the $\Delta (1232)$, $N (1535)$, and $N
(1650)$ resonances, since there are several implications that these
hadrons may have certain fractions of the meson--baryon components.
For $\Delta (1232)$, there are several suggestions that the effect of
the meson cloud seems to be large, for instance, in the $M1$
transition form factor for $\gamma ^{\ast} N \to \Delta (1232)$ at
$Q^{2} = 0$~\cite{Sato:2009de}.  The $\pi N$ compositeness for $\Delta
(1232)$ has been already studied in a simple phenomenological
model~\cite{Aceti:2014ala}, implying large contribution of the $\pi N$
component to the internal structure of $\Delta (1232)$.  On the other
hand, for $N (1535)$ and $N (1650)$, there are several studies that
they can be dynamically generated from meson--baryon degrees of
freedom without introducing explicit resonance poles in the so-called
chiral unitary approach~\cite{Kaiser:1995cy, Nieves:2001wt,
  Inoue:2001ip, Hyodo:2002pk, Hyodo:2003qa, Bruns:2010sv,
  Khemchandani:2013nma, Garzon:2014ida}.  In this approach, the $\pi
N$ and its coupled-channels amplitude is obtained based on the
combination of chiral perturbation theory and the unitarization of the
scattering amplitude~\cite{Kaiser:1995cy, Meissner:1999vr,
  Nieves:2001wt, Inoue:2001ip, Hyodo:2002pk, Hyodo:2003qa,
  Bruns:2010sv, Alarcon:2012kn, Khemchandani:2013nma, Garzon:2014ida}.
The results in the chiral unitary approach might suggest that $N
(1535)$ and $N (1650)$ are meson--baryon molecular states.  Of special
interest is the relation between $N (1535)$, $N (1650)$, and other
dynamically generated resonances in the chiral unitary approach such
as $\Lambda (1405)$ and $\Xi (1690)$.  Namely, it is suggested in
Ref.~\cite{GarciaRecio:2003ks} that, in a flavor SU(3) symmetric world
in the chiral unitary approach, $N (1535)$ and $N (1650)$ degenerate,
together with the one of the two-$\Lambda (1405)$ pole, $\Xi (1690)$,
and so on, into two degenerated octets as dynamically generated
states.  Since both $\Lambda (1405)$ and $\Xi (1690)$ in the chiral
unitary approach in the physical world are respectively found to be
indeed the $\bar{K} N$~\cite{Sekihara:2014kya} and $\bar{K}
\Sigma$~\cite{Sekihara:2015qqa} molecular states in terms of the
compositeness, the degeneracy in the flavor SU(3) symmetric world
implies that both $N (1535)$ and $N (1650)$ may be meson--baryon
molecular states as well.  However, the compositeness for $N (1535)$
was studied in the chiral unitary approach with the simplest
interaction, i.e., the Weinberg--Tomozawa term, in
Ref.~\cite{Sekihara:2014kya}, and the result indicated the large
component originating from contributions other than the pseudoscalar
meson--baryon dynamics considered for $N (1535)$.

Motivated by these observations, in the present study, we evaluate the
compositeness for the $\Delta (1232)$, $N (1535)$, and $N (1650)$
resonances and $N (940)$ from the precise $\pi N$ scattering
amplitudes in the chiral unitary approach, taking into account the
interaction up to the next-to-leading order from chiral perturbation
theory and including an explicit $\Delta$ term for $\Delta (1232)$.
For $\Delta (1232)$, we discuss its $\pi N$ compositeness in the $\pi
N$ single-channel scattering, while we treat $N (1535)$ and $N (1650)$
in a $\pi N$-$\eta N$-$K \Lambda$-$K \Sigma$ coupled-channels problem
without introducing explicit bare states.  The loop function in our
approach is evaluated with the dimensional regularization.  We fit the
model parameters to the solution of the partial wave analysis for the
$\pi N$ scattering amplitude, and calculate the meson--baryon
compositeness for $\Delta (1232)$, $N(1535)$, and $N(1650)$ from the
$\pi N$ scattering amplitude.  A part of the study on $\Delta (1232)$
was already reported in Ref.~\cite{Sekihara:2015aba}.  In this paper,
we show the details of the formulation, results, and discussions.

This paper is organized as follows.  In Sec.~\ref{sec:2} we formulate
the compositeness for baryonic resonances in order to discuss the
meson--baryon molecular structure inside the resonances.  In the
formulation we will consider the case of a relativistic scattering of
arbitrary spin particles.  Next, in Sec.~\ref{sec:3} we shown our
numerical calculations on the compositeness for the $\Delta (1232)$,
$N (1535)$, and $N (1650)$ resonances in the chiral unitary approach
with the interaction up to the next-to-leading order in chiral
perturbation theory.  Section~\ref{sec:4} is devoted to the summary of
this study and outlook.


\section{Compositeness}
\label{sec:2}

First of all we formulate the compositeness, which has been recently
developed in the hadron physics so as to discuss the hadronic
molecular components inside hadrons.  The compositeness is defined as
contributions from two-body wave functions to the normalization of the
total wave function $| \Psi \rangle$ for the resonance state, and
corresponds to unity minus the field renormalization constant
intensively discussed in the 1960s~\cite{Salam:1962ap,
  Weinberg:1962hj, Ezawa:1963zz, Lurie:1963}.  Although the
compositeness is not observable and hence a model dependent quantity,
it will be an important piece of information on the internal structure
of the resonance state.

In this section we first show how to extract the compositeness from
the residue of the two-body to two-body scattering amplitude at the
resonance pole in the nonrelativistic framework in Sec.~\ref{sec:2A}.
Next we extend our discussions to the relativistic case in
Sec.~\ref{sec:2B}.  Both in Sec.~\ref{sec:2A} and Sec.~\ref{sec:2B} we
do not specify the form of the interaction so as to give the general
formulae of the compositeness in terms of the residue of the
scattering amplitude at the resonance pole position, and in
Sec.~\ref{sec:2C} we consider the formulation with the separable
interaction, which is employed in our numerical calculations in
Sec.~\ref{sec:3}.  Then we give several comments on the interpretation
of the compositeness for resonance states in Sec.~\ref{sec:2D}.  In
the following we take the rest frame of the center-of-mass motion,
namely two scattering particles have equal and opposite momentum, and
hence the resonance state is at rest with zero momentum.

\subsection{Scattering amplitude and wave function \\* in a
  nonrelativistic case}
\label{sec:2A}

We consider a two-body to two-body coupled-channels scattering in a
nonrelativistic condition governed by the interaction operator
$\hat{V}$ for the two-body systems, with which we have only the
two-body states in the practical model space.  For simplicity, we
assume that the interaction is a central force and neglect the spin of
the scattering particles.  Moreover, for the later applications we
allow the interaction to depend intrinsically on the energy of the
system $E$, which corresponds to the eigenenergy of the full
Hamiltonian.\footnote{If there are missing (or implicit) channels
  which are implemented in the interaction, such missing channels,
  regardless of one-body bare states or more than one-body scattering
  states of higher thresholds, can be origin of the intrinsic energy
  dependence of the interaction.  See Ref.~\cite{Sekihara:2014kya} for
  the details.}  The scattering amplitude can be formally obtained
with the Lippmann--Schwinger equation in an operator form:
\begin{align}
\hat{T} ( E ) = & \hat{V} ( E ) 
+ \hat{V} ( E ) \frac{1}{E - \hat{H}_{0}} \hat{T} ( E ) 
\notag \\
= & \hat{V} ( E ) + \hat{V} ( E ) \frac{1}{E - \hat{H}} \hat{V} ( E ) ,
\label{eq:LS}
\end{align}
with the $T$-matrix operator $\hat{T}$, the free Hamiltonian
$\hat{H}_{0}$, and the full Hamiltonian $\hat{H} \equiv \hat{H}_{0} +
\hat{V} ( E )$.

First, in order to evaluate the scattering amplitude from the
Lippmann--Schwinger equation~\eqref{eq:LS}, we have to introduce the
scattering states with which we calculate the matrix element of the
$T$-matrix operator.  We represent the $j$th channel two-body
scattering state with relative momentum $\bm{q}$ as $| \bm{q}_{j}
\rangle$, which is an eigenstate of the free Hamiltonian
$\hat{H}_{0}$:
\begin{equation}
\hat{H}_{0} | \bm{q}_{j} \rangle = 
E_{j} ( q ) | \bm{q}_{j} \rangle , 
\quad 
\langle \bm{q}_{j} | \hat{H}_{0} = 
E_{j} ( q ) \langle \bm{q}_{j} | ,
\end{equation}
where $q \equiv | \bm{q} |$ is the magnitude of the momentum $\bm{q}$
and the eigenenergy $E_{j} ( q )$ contains the threshold energy:
\begin{equation}
E_{j} ( q ) 
\equiv m_{j} + M_{j} + \frac{q^{2}}{2 \mu _{j}} ,
\quad 
\mu _{j} \equiv \frac{m_{j} M_{j}}{m_{j} + M_{j}} ,
\label{eq:Ej_NR}
\end{equation}
with the masses of the $j$th channel particles $m_{j}$ and $M_{j}$.
We fix the normalization of the scattering states as
\begin{equation}
\langle \bm{q}_{k}^{\prime} | \bm{q}_{j} \rangle 
= ( 2 \pi )^{3} \delta _{j k} \delta ( \bm{q}^{\prime} - \bm{q} ) .
\end{equation}

Now we can express scattering amplitude of the $k ( \bm{q} ) \to j (
\bm{q}^{\prime} )$ scattering, where $\bm{q}^{( \prime )}$ is the
relative momenta in the initial (final) state, as
\begin{equation}
\langle \bm{q}_{j}^{\prime} | \hat{T} ( E ) | \bm{q}_{k} \rangle 
\equiv T_{j k} ( E  ; \, \bm{q}^{\prime} , \, \bm{q} ) ,
\end{equation}
which is obtained from the interaction 
\begin{equation}
\langle \bm{q}_{j}^{\prime} | \hat{V} ( E ) | \bm{q}_{k} \rangle 
\equiv V_{j k} ( E  ; \, \bm{q}^{\prime} , \, \bm{q} ) .
\end{equation}
In this study we assume the time-reversal invariance of the scattering
process.  This constrains the interaction and amplitude, with an
appropriate choice of phases of the states, as
\begin{equation}
V_{j k} ( E  ; \, \bm{q}^{\prime} , \, \bm{q} ) 
= V_{k j} ( E  ; \, \bm{q} , \, \bm{q}^{\prime} ) , 
\end{equation}
\begin{equation}
T_{j k} ( E  ; \, \bm{q}^{\prime} , \, \bm{q} ) 
= T_{k j} ( E  ; \, \bm{q} , \, \bm{q}^{\prime} ) . 
\end{equation}
The scattering amplitude $T_{j k} ( E; \, \bm{q}^{\prime} , \,
\bm{q})$ is a solution of the Lippmann--Schwinger equation in the
following form:
\begin{align}
& T_{j k} ( E  ; \, \bm{q}^{\prime} , \, \bm{q} ) 
= V_{j k} ( E  ; \, \bm{q}^{\prime} , \, \bm{q} ) 
\notag \\
& + \sum _{l} \int \frac{d^{3} k}{( 2 \pi )^{3}} 
\frac{V_{j l} ( E  ; \, \bm{q}^{\prime} , \, \bm{k} ) 
T_{l k} ( E  ; \, \bm{k} , \, \bm{q} )}
{E - E_{l} ( k )} .
\label{eq:LS-II}
\end{align}
In the actual scattering, the system in the initial and final states
should be on mass shell and the energy should be determined as $E =
E_{j} ( q^{\prime} ) = E_{k} ( q )$.  We call this scattering
amplitude as the on-shell amplitude.  However, in the intermediate
state the energy $E_{l} ( k )$ takes different values from $E$.
Moreover, we can mathematically perform the analytic continuation of
the scattering amplitude by taking the value of the energy $E$
different from $E_{j} ( q^{\prime} ) = E_{k} ( q )$ as an off-shell
amplitude.  This will be essential to extract the wave function from
the scattering amplitude at the resonance pole position in the complex
energy plane.

Next, it is useful to decompose the scattering amplitude into partial
wave amplitudes:
\begin{equation}
T_{j k} ( E  ; \, \bm{q}^{\prime} , \, \bm{q} ) 
= \sum _{L = 0}^{\infty} ( 2 L + 1 ) 
T_{L, j k} ( E  ; \, q^{\prime} , \, q ) 
P_{L} ( \hat{q}^{\prime} \cdot \hat{q} ) ,
\end{equation}
and in a similar manner for the interaction $V$, where $P_{L}$ is the
Legendre polynomials and $\hat{q}^{( \prime )}$ is the unit vector for
the direction of $\bm{q}^{( \prime )}$: $\hat{q}^{( \prime )} \equiv
\bm{q}^{( \prime )} / q^{( \prime )}$.  Each partial wave amplitude can
be extracted as
\begin{equation}
T_{L, j k} ( E  ; \, q^{\prime} , \, q ) 
= \frac{1}{2} \int _{-1}^{1} d ( \hat{q}^{\prime} \cdot \hat{q} )
P_{L} ( \hat{q}^{\prime} \cdot \hat{q} ) 
T_{j k} ( E  ; \, \bm{q}^{\prime} , \, \bm{q} ) .
\label{eq:PWA_dec}
\end{equation}
Since the Legendre polynomials satisfy the following relation
\begin{equation}
\int d \Omega _{\bm{k}} P_{L} ( \hat{q}^{\prime} \cdot \hat{k} )
P_{L^{\prime}} ( \hat{k} \cdot \hat{q} )
= \frac{4 \pi}{2 L + 1} \delta _{L L^{\prime}} 
P_{L} ( \hat{q}^{\prime} \cdot \hat{q} ) ,
\end{equation}
for the integral with respect to the solid angle of a vector $\bm{k}$,
$\Omega _{\bm{k}}$, we can rewrite the Lippmann--Schwinger
equation~\eqref{eq:LS-II} as
\begin{align}
& T_{L, j k} ( E  ; \, q^{\prime} , \, q ) 
= V_{L, j k} ( E  ; \, q^{\prime} , \, q ) 
\notag \\
& + \sum _{l} \int _{0}^{\infty} \frac{d k}{2 \pi ^{2}} k^{2}
\frac{V_{L, j l} ( E  ; \, q^{\prime} , \, k ) 
T_{L, l k} ( E  ; \, k , \, q )}
{E - E_{l} ( k )} .
\label{eq:LS-III}
\end{align}
We note that in our formulation the on-shell scattering amplitude in
each partial wave satisfies the optical theorem from the unitarity of
the $S$-matrix in the following normalization:
\begin{equation}
\text{Im} \, T_{L, j j}^{\text{on-shell}} ( E )
= - \sum _{k} \frac{\mu _{k} q_{k}}{2 \pi} 
\left | T_{L, j k}^{\text{on-shell}} ( E )
\right | ^{2} ,
\end{equation}
where $q_{k} \equiv \sqrt{2 \mu _{k} (E - m_{k} - M_{k})}$ is the
on-shell relative momentum in the $k$th channel and the sum runs over
the open channels.

Let us now suppose that there is a resonance state $| \psi _{L M}
\rangle$ in the partial wave $L$ with its azimuthal component $M$.
Here, in order to ensure a finite normalization of the resonance wave
function $| \psi _{L M} \rangle$, we employ the Gamow vector, which
was first introduced to describe unstable nuclei~\cite{Hokkyo:1965zz,
  Berggren:1968zz, Romo:1968zz, Hernandez:1984zzb}.  The resonance
state $| \psi _{L M} \rangle$ as the Gamow vector is a solution of the
\Schr equation:
\begin{equation}
\hat{H} | \psi _{L M} \rangle 
= \left [ \hat{H}_{0} + \hat{V} ( E_{\rm pole} ) \right ] | \psi _{L M} \rangle
= E_{\rm pole} | \psi _{L M} \rangle , 
\end{equation}
with the eigenenergy $E_{\rm pole}$.  We note that the resonance
eigenenergy is in general complex, $E_{\rm pole}^{\ast} \ne E_{\rm
  pole}$; $\text{Re} E_{\rm pole}$ and $- 2 \, \text{Im} E_{\rm pole}$
are the mass and width of the resonance state, respectively.  Then, to
establish the normalization of the resonance state as the Gamow
vector, we take $\langle \psi _{L M}^{\ast} |$ instead of $\langle
\psi _{L M} |$ for the bra vector of the resonance.  In this notation
we can normalize the resonance wave function in the following manner:
\begin{equation}
\langle \psi _{L M^{\prime}}^{\ast} | \psi _{L M} \rangle 
= \delta _{M^{\prime} M} .
\end{equation}
The \Schr equation for the resonance bra state is expressed with the
same eigenenergy as
\begin{equation}
\langle \psi _{L M}^{\ast} | \hat{H} 
= \langle \psi _{L M}^{\ast} 
| \left [ \hat{H}_{0} + \hat{V} ( E_{\rm pole} ) \right ]
= \langle \psi _{L M}^{\ast} | E_{\rm pole} .
\end{equation}
Here we summarize the two-body component of the resonance wave
function in momentum space.  Namely, since the interaction is assumed
to be a central force, for the $L$-wave resonance, the wave function
in momentum space can be written as a product of the radial part
$R_{j} ( q )$ in $j$ channel and the spherical harmonics $Y_{L M} (
\hat{q} )$ as
\begin{equation}
\langle \bm{q}_{j} | \psi _{L M} \rangle 
= R_{j} ( q ) Y_{L M} ( \hat{q} ) .
\label{eq:psiq}
\end{equation}
We fix the normalization of the spherical harmonics $Y_{L M} ( \hat{q}
)$ as
\begin{equation}
\int d \Omega _{\bm{q}} Y_{L M} ( \hat{q} )
Y_{L^{\prime} M^{\prime}}^{\ast} ( \hat{q} ) 
= 4 \pi \delta _{L L^{\prime}} \delta _{M M^{\prime}}.
\end{equation}
On the other hand, from the bra state $\langle \psi _{L M}^{\ast} |$
the two-body wave function can be evaluated as
\begin{equation}
\langle \psi _{L M}^{\ast} | \bm{q}_{j} \rangle 
= R_{j} ( q ) Y_{L M}^{\ast} ( \hat{q} ) .
\label{eq:psiq-ast}
\end{equation}
Here we emphasize that, while we take the complex conjugate for the
spherical harmonics, we do not take for the radial part.  This is
because, while the spherical part can be calculated and normalized in
a usual sense, the radial part should be treated so as to remove the
divergence of the wave function at $q \to \infty$~\cite{Hokkyo:1965zz,
  Berggren:1968zz, Romo:1968zz, Hernandez:1984zzb} when we calculate
the norm.  From the above wave function, we can calculate the norm
with respect to the $j$th channel two-body wave function, $X_{j}$, in
the following manner:
\begin{equation}
X_{j} \equiv \int \frac{d^{3} q}{( 2 \pi )^{3}} 
\langle \psi _{L M}^{\ast} | \bm{q}_{j} \rangle 
\langle \bm{q}_{j} | \psi _{L M} \rangle 
= \int _{0}^{\infty} \frac{d q}{2 \pi ^{2}} q^{2}
\left [ R_{j} ( q ) \right ] ^{2} .
\label{eq:norm}
\end{equation}
This quantity is referred to as the compositeness.  In this
construction, the compositeness $X_{j}$ is given by the complex number
squared of the radial part $R_{j} ( q )$ rather than by the absolute
value squared, which is essential to normalize the resonance wave
function.  Therefore, in general the compositeness becomes complex for
resonance states.  We also note that the sum of the norm $X_{j}$
should be unity if there is no missing channels, which would be an
eigenstate of the free Hamiltonian, to describe the resonance state.
However, in actual calculations we may have contributions from missing
channels, which can be implemented as the energy dependence of the
interaction.  As we will discuss when we introduce the scattering
amplitude and its residue at the resonance pole position, we do not
make the sum of the compositeness $X_{j}$ coincide with unity by hand.
Instead, the value of the norm is automatically fixed when we
calculate the residue of the scattering amplitude.  Here we
representatively denote the missing channels as $| \psi _{0} \rangle$,
which represents not only one-body bare states but also more than
one-body scattering states.  In the present notation we can decompose
unity in terms of the eigenstates of the free Hamiltonian:
\begin{equation}
\one = | \psi _{0} \rangle \langle \psi _{0} | 
+ \sum _{j} \int \frac{d^{3} q}{( 2 \pi )^{3}} 
| \bm{q}_{j} \rangle \langle \bm{q}_{j} | .
\end{equation}
Therefore, the normalization of the resonance wave function $| \psi
_{L M} \rangle$ is expressed as 
\begin{equation}
\langle \psi _{L M}^{\ast} | \psi _{L M} \rangle 
= Z + \sum _{j} X_{j} = 1 ,
\end{equation}
where we have introduced the missing-channel contributions $Z$
defined as
\begin{equation}
Z \equiv  \langle \psi _{L M}^{\ast} 
| \psi _{0} \rangle \langle \psi _{0} | \psi _{L M} \rangle . 
\end{equation}
Note that the quantity $Z$, which has been referred to as the
elementariness\footnote{Although $Z$ is called elementariness, it
  contains contributions not only from elementary one-body states but
  also from more than one-body scattering states.}, becomes complex
for resonance states as well.  The explicit form of the elementariness
$Z$ will be given in Sec.~\ref{sec:2C} in our model.

We now establish the way to extract the compositeness from the
off-shell scattering amplitude obtained by the analytic continuation
for the energy.  The key is the fact that the resonance wave function
appears as the residue at the resonance pole of the scattering
amplitude.  Namely, near the resonance pole, the off-shell scattering
amplitude is dominated by the resonance pole term in the expansion by
the eigenstates of the full Hamiltonian, and hence we have [see the
  last expression in Eq.~\eqref{eq:LS}]
\begin{equation}
\hat{T} ( E ) \approx \sum _{M = - L}^{L} 
\hat{V} ( E_{\rm pole} ) | \psi _{L M} \rangle \frac{1}{E - E_{\rm pole}}
\langle \psi _{L M}^{\ast} | \hat{V} ( E_{\rm pole} ) ,
\label{eq:Tapprox}
\end{equation}
where we have summed up the possible azimuthal component $M$.
Calculating the matrix element of this $T$-matrix operator, we obtain
\begin{align}
& T_{j k} ( E ; \, \bm{q}^{\prime} , \, \bm{q} ) 
\notag \\
& \approx \sum _{M = - L}^{L} 
\frac{\langle \bm{q}_{j}^{\prime} | \hat{V} ( E_{\rm pole} )
  | \psi _{L M} \rangle 
  \langle \psi _{L M}^{\ast} | \hat{V} ( E_{\rm pole} ) | \bm{q}_{k} \rangle}
     {E - E_{\rm pole}} .
\end{align}
Then we need to evaluate the matrix elements in the numerator,
$\langle \bm{q}_{j}^{\prime} | \hat{V} ( E_{\rm pole} ) | \psi _{L M}
\rangle$ and $\langle \psi _{L M}^{\ast} | \hat{V} ( E_{\rm pole} ) |
\bm{q}_{k} \rangle$.  We can evaluate the former one by using the
\Schr equation as
\begin{align}
\langle \bm{q}_{j} | \hat{V} ( E_{\rm pole} ) | \psi _{L M} \rangle
= & \langle \bm{q}_{j} | \left ( \hat{H} - \hat{H}_{0} \right ) 
| \psi _{L M} \rangle 
\notag \\
= & \left [ E_{\rm pole} - E_{j} ( q ) \right ] 
\langle \bm{q}_{j} | \psi _{L M} \rangle , 
\end{align}
and from Eq.~\eqref{eq:psiq} we obtain
\begin{equation}
\langle \bm{q}_{j} | \hat{V} ( E_{\rm pole} ) | \psi _{L M} \rangle
= \gamma _{j} ( q ) Y_{L M} ( \hat{q} ) , 
\end{equation}
with
\begin{equation}
\gamma _{j} ( q ) \equiv 
\left [ E_{\rm pole} - E_{j} ( q ) \right ] 
R_{j} ( q ) .
\label{eq:gamma}
\end{equation}
In a similar manner we can calculate the latter matrix element as
\begin{equation}
\langle \psi _{L M}^{\ast} | \hat{V} ( E_{\rm pole} ) | \bm{q}_{j} \rangle
= \gamma _{j} ( q ) Y_{L M}^{\ast} ( \hat{q} ) .
\end{equation}
By using the above matrix elements, we can rewrite the scattering
amplitude near the resonance pole as
\begin{align}
T_{j k} ( E ; \, \bm{q}^{\prime} , \, \bm{q} ) 
& \approx 
\frac{\gamma _{j} ( q^{\prime} ) \gamma _{k} ( q )}{E - E_{\rm pole}} 
\sum _{M = - L}^{L} 
Y_{L M} ( \hat{q}^{\prime} ) Y_{L M}^{\ast} ( \hat{q} ) 
\notag \\
& = ( 2 L + 1 )
\frac{\gamma _{j} ( q^{\prime} ) \gamma _{k} ( q )}{E - E_{\rm pole}} 
P_{L} ( \hat{q}^{\prime} \cdot \hat{q} ) ,
\label{eq:Tapprox-II}
\end{align}
where we have used the formula for the spherical harmonics and
Legendre polynomials:
\begin{equation}
\sum _{M = - L}^{L} 
Y_{L M} ( \hat{q}^{\prime} ) Y_{L M}^{\ast} ( \hat{q} ) 
= (2 L + 1) P_{L} ( \hat{q}^{\prime} \cdot \hat{q} ) .
\end{equation}
The expression in Eq.~\eqref{eq:Tapprox-II} indicates that the partial
wave amplitude in $L$ wave contains the resonance pole, as we
expected:
\begin{equation}
T_{L, j k} ( E ; \, q^{\prime} , \, q ) = 
\frac{\gamma _{j} ( q^{\prime} ) \gamma _{k} ( q )}{E - E_{\rm pole}}
+ (\text{regular at } E = E_{\rm pole}) .
\end{equation}
Furthermore, the residue of the partial wave amplitude contains
information on the resonance wave function via the expression in
Eq.~\eqref{eq:gamma}.  Actually, we can calculate the $j$th channel
compositeness, $X_{j}$, by using the residue $\gamma _{j} ( q )$ as
\begin{align}
X_{j} = & \int _{0}^{\infty} \frac{d q}{2 \pi ^{2}} q^{2}
\left [ R_{j} ( q ) \right ] ^{2}
\notag \\
= & \int _{0}^{\infty} \frac{d q}{2 \pi ^{2}} q^{2}
\left [ \frac{\gamma _{j} ( q )}
{E_{\rm pole} - E_{j} ( q )} \right ] ^{2} .
\label{eq:X_gamma}
\end{align}
This is the formula to evaluate the $j$th channel compositeness
$X_{j}$ from the residue of the partial wave amplitude $T_{L}$ at the
resonance pole.  An important point is that the residue $\gamma _{j} (
q )$ is obtained from the Lippmann--Schwinger equation without
introducing any extra factor to scale the value of the compositeness
$X_{j}$.  In this sense, the value of the norm in Eq.~\eqref{eq:norm}
is automatically fixed when we calculate the residue of the scattering
amplitude. Indeed, it was proved in Ref.~\cite{Hernandez:1984zzb} that
the wave function from the residue of the scattering amplitude is
correctly normalized to be unity for a general energy independent
interaction.\footnote{An analysis on the relation between the residue
  of the scattering amplitude and the wave function will be presented
  in detail elsewhere~\cite{Sekihara:2015b}. }

Here we note that the compositeness $X_{j}$ is not observable and
hence in general a model dependent quantity.  This can be understood
with the property of the residue $\gamma _{j} ( q )$.  Namely, while
the on-shell scattering amplitude for open channels is observable, the
off-shell amplitude with the energy analytically continued to the
resonance pole position is not observable.  Therefore, in order to
calculate the residue $\gamma _{j} ( q )$, in general one needs some
model or assumptions for the analytic continuation.  In other words,
we have to fix the functional form when we evaluate the off-shell
scattering amplitude.  This is reflected as the model dependence of
the residue $\gamma _{j}$ and hence the compositeness $X_{j}$.
However, in certain cases we can express the compositeness only with
the observable quantities.  A special case is that the pole exists
very close to the on-shell energies, in which we can directly relate
the compositeness with threshold parameters such as the scattering
length and effective range~\cite{Weinberg:1965zz, Baru:2003qq,
  Hyodo:2013iga, Hanhart:2014ssa, Hyodo:2014bda, Kamiya:2015aea}.

Finally we comment on the semirelativistic case, in which the
eigenenergy of the free Hamiltonian~\eqref{eq:Ej_NR} is replaced with
\begin{equation}
E_{j} ( q ) 
\equiv \sqrt{q^{2} + m_{j}^{2}} + \sqrt{q^{2} + M_{j}^{2}} .
\label{eq:Ej_SR}
\end{equation}
Even in this case we can follow the same discussion, and we obtain the
same formula for the compositeness~\eqref{eq:X_gamma} but with the
semirelativistic eigenenergy $E_{j} ( q )$ in Eq.~\eqref{eq:Ej_SR}.

\subsection{Scattering amplitude and wave function \\* in a
  relativistic case}
\label{sec:2B}

We extend our discussions to the relativistic case of the two-body to
two-body scattering $k ( p^{\mu}, \, q^{\mu} ) \to j ( p^{\prime \mu},
\, q^{\prime \mu} )$, where $j$ and $k$ are channel indices and
$p^{\mu}$, $q^{\mu}$, $p^{\prime \mu}$, and $q^{\prime \mu}$ are the
momenta of particles whose masses are $M_{k}$, $m_{k}$, $M_{j}$, and
$m_{j}$, respectively.  We first consider a scattering of two spinless
particles, and then we treat a two-body scattering of arbitrary spins
by using the partial wave amplitude.  We take the center-of-mass
frame, where the total energy-momentum of the system becomes $P^{\mu}
= p^{\mu} + q^{\mu} = p^{\prime \mu} + q^{\prime \mu} = ( w , \,
\bm{0})$ with the center-of-mass energy $w$.  The conventions used in
this study is summarized in Appendix~\ref{app:1}.

In general, the scattering amplitude of two spinless particles is
expressed as a function of the Mandelstam variable $s \equiv w^{2}$
and momenta $q^{\mu}$ and $q^{\prime \mu}$.  The scattering amplitude
is a solution of the Lippmann--Schwinger equation in a relativistic
form:
\begin{align}
& T_{j k} ( s; \, q^{\prime \mu}, \, q^{\mu} )
= V_{j k} ( s; \, q^{\prime \mu}, \, q^{\mu} )
\notag \\
& + i \sum _{l} \int \frac{d^{4} k}{( 2 \pi )^{4}} 
\frac{V_{j l} ( s; \, q^{\prime \mu}, \, k^{\mu} )
T_{l k} ( s; \, k^{\mu}, \, q^{\mu} ) }
     {( k^{\mu} k_{\mu} - m_{j}^{2} )
       [ ( P - k )^{\mu} ( P - k )_{\mu} - M_{j}^{2} ]} .
\label{eq:LS-IV}
\end{align}
Here we allow that the interaction kernel $V_{j k} ( s; \, q^{\prime
  \mu}, \, q^{\mu} )$ may contain, in addition to the tree-level
parts, contributions from $t$- and $u$-channel loops in a usual manner
of the quantum field theory.  For the on-shell amplitude $s$ is
related to the momenta $q^{\mu}$ and $q^{\prime \mu}$, while the
off-shell amplitude can be obtained with the analytic continuation to
the complex values of $s$.

From the Lippmann--Schwinger equation~\eqref{eq:LS-IV}, we construct
an analogue to the scattering equation in the nonrelativistic case,
which will be essential to relate the scattering amplitude with the
wave function clearly.  To this end, we assume an on-shell condition
to the energy $q^{( \prime ) 0}$ inside $V_{j k} ( s; \, q^{\prime
  \mu} , \, q^{\mu} )$ by making it a function of the center-of-mass
energy $w$ as
\begin{equation}
\begin{split}
  & q^{0} \to \omega _{k} ( s ) \equiv
  \frac{s + m_{k}^{2} - M_{k}^{2}}{2 \sqrt{s}} ,
  \\
  & q^{\prime 0} \to \omega _{j}^{\prime} ( s ) 
  \equiv \frac{s + m_{j}^{2} - M_{j}^{2}}{2 \sqrt{s}} .
\end{split}
\label{eq:q0_onshell}
\end{equation}
In this assumption, we can treat the interaction kernel as in the
nonrelativistic form:
\begin{equation}
V_{j k} ( s ; \, q^{\prime \mu} , \, q^{\mu} ) 
\to V_{j k} ( s ; \, \bm{q}^{\prime} , \, \bm{q} ) ,
\end{equation}
and hence, after performing the $k^{0}$ integral, the
Lippmann--Schwinger equation~\eqref{eq:LS-IV} becomes
\begin{align}
& T_{j k} ( s, \, \bm{q}^{\prime}, \, \bm{q} )
= V_{j k} ( s, \, \bm{q}^{\prime}, \, \bm{q} )
\notag \\
& + \sum _{l} \int \frac{d^{3} k}{( 2 \pi )^{3}} 
\frac{\sqrt{s_{l} ( k )}} 
{2 \omega _{l} ( k ) \Omega _{l} ( k )}
\frac{V_{j l} ( s, \, \bm{q}^{\prime}, \, \bm{k} )
T_{l k} ( s, \, \bm{k}, \, \bm{q} ) }
{s - s_{l} ( k )} ,
\label{eq:LS-V}
\end{align}
with
\begin{equation}
\omega _{j} ( q ) \equiv \sqrt{q^{2} + m_{j}^{2}} , 
\quad 
\Omega _{j} ( q ) \equiv \sqrt{q^{2} + M_{j}^{2}} ,
\end{equation}
\begin{equation}
s_{j} ( q ) \equiv [ \omega _{j} ( q ) + \Omega _{j} ( q ) ]^{2} .
\end{equation}
Now we can perform the partial wave decomposition in a similar manner
as in the nonrelativistic case.  In particular, in the present
formulation the partial wave amplitude $T_{L, \, j k}$, extracted in
the same way as in Eq.~\eqref{eq:PWA_dec}, is the solution of the
Lippmann--Schwinger equation
\begin{align}
& T_{L, \, j k} ( s; \, q^{\prime}, \, q )
= V_{L, \, j k} ( s; \, q^{\prime}, \, q )
\notag \\
& + \sum _{l} \int \frac{d k}{2 \pi ^{2}} 
\frac{k^{2} \sqrt{s_{l} ( k )}}
{2 \omega _{l} ( k ) \Omega _{l} ( k )}
\frac{V_{L, \, j l} ( s, \, q^{\prime}, \, k )
T_{L, \, l k} ( s, \, k, \, q ) }
{s - s_{l} ( k )} ,
\label{eq:LS-VI}
\end{align}
and satisfies the optical theorem
\begin{equation}
\text{Im} \, T_{L, \, j j}^{\text{on-shell}} ( s )
= - \sum _{k} \frac{q_{k} ( s )}{8 \pi \sqrt{s}} 
\left | T_{L, j k}^{\text{on-shell}} ( s ) \right | ^{2} ,
\label{eq:opt-rel}
\end{equation}
where $q_{k} ( s )$ is the $k$th channel center-of-mass momentum in
the relativistic form
\begin{equation}
q_{k} ( s ) \equiv 
\frac{\lambda ^{1/2} ( s , \, m_{k}^{2} , \, M_{k}^{2} )}
{2 \sqrt{s}} ,
\end{equation}
with the \Kaellen function $\lambda (x, \, y, \, z) = x^{2} + y^{2} +
z^{2} - 2 x y - 2 y z - 2 z x$, and the sum runs over the open
channels.

Of special interest is the expression of the Lippmann--Schwinger
equation in Eq.~\eqref{eq:LS-V}, with which we can apply the
relativistic formulation of the wave function developed in
Refs.~\cite{Sekihara:2014kya, Mandelzweig:1986hk, Wallace:1989nm}.  In
this formulation, the two-body equation for the resonance state $|
\Psi _{L M} \rangle$, in the partial wave $L$ with its azimuthal
component $M$, is expressed in an extended form of the \Schr equation
as~\cite{Sekihara:2014kya}
\begin{equation}
\begin{split}
& \left [ \hat{\mathcal{K}} + \hat{\mathcal{V}} ( s_{\rm pole} ) \right ]
| \Psi _{L M} \rangle 
= s_{\rm pole} | \Psi _{L M} \rangle , 
\\
& \langle \Psi _{L M}^{\ast} |
\left [ \hat{\mathcal{K}} + \hat{\mathcal{V}} ( s_{\rm pole} ) \right ]
= s_{\rm pole} \langle \Psi _{L M}^{\ast} | ,
\end{split}
\label{eq:Psi_rel}
\end{equation}
where $\hat{\mathcal{K}}$ and $\hat{\mathcal{V}}$ are the kinetic
energy and interaction operators, respectively, and $s_{\rm pole}$ is
the resonance pole position with respect to the Mandelstam variable
$s$.  The kinetic operator $\hat{\mathcal{K}}$ corresponds to the free
Hamiltonian in the nonrelativistic framework and has eigenstates of
the $j$th channel two-body covariant scattering state with the
relative momentum $\bm{q}$, $| \bm{q}_{j} \rangle _{\text{co}}$, with
which eigenvalues of the kinetic operator are
\begin{equation}
\hat{\mathcal{K}} | \bm{q}_{j} \rangle _{\text{co}}
= s_{j} ( q ) | \bm{q}_{j} \rangle _{\text{co}},
\quad 
{}_{\text{co}} \langle \bm{q}_{j} | \hat{\mathcal{K}} 
= s_{j} ( q ) \; {}_{\text{co}} \langle \bm{q}_{j} | .
\end{equation}
In this study we take the same normalization of the covariant
scattering state as in Ref.~\cite{Sekihara:2014kya}:
\begin{equation}
{}_{\text{co}} \langle \bm{q}_{k}^{\prime} | \bm{q}_{j} \rangle _{\text{co}} 
= \frac{2 \omega _{j} ( q ) \Omega _{j} ( q )}{\sqrt{s_{j} ( q )}}
( 2 \pi )^{3} \delta _{j k} \delta ( \bm{q}^{\prime} - \bm{q} ) .
\label{eq:norm_qco}
\end{equation}
The factor $2 \omega _{j} ( q ) \Omega _{j} ( q ) / \sqrt{s_{j} ( q
  )}$ guarantees that the measure of the integral in the expression of
the compositeness is Lorentz invariant, as we will see later.  The
wave function in momentum space is expressed as
\begin{equation}
\begin{split}
& {}_{\text{co}} \langle \bm{q}_{j} | \Psi _{L M} \rangle 
= R_{j} ( q ) Y_{L M} ( \hat{q} ) ,
\\
& \langle \Psi _{L M}^{\ast} | \bm{q}_{j} \rangle _{\text{co}}
= R_{j} ( q ) Y_{L M}^{\ast} ( \hat{q} ) .
\end{split}
\end{equation}
With this scattering state, we can calculate the interaction kernel
$V_{j k}$ in the following manner:
\begin{equation}
{}_{\text{co}} \langle \bm{q}_{j}^{\prime} | 
\hat{\mathcal{V}} ( s )
| \bm{q}_{k} \rangle _{\text{co}}
= V_{j k} ( s ; \, \bm{q}^{\prime}, \, \bm{q} ) ,
\end{equation}
and similarly the scattering amplitude is calculated as 
\begin{equation}
{}_{\text{co}} \langle \bm{q}_{j}^{\prime} | 
\hat{\mathcal{T}} ( s )
| \bm{q}_{k} \rangle _{\text{co}}
= T_{j k} ( s ; \, \bm{q}^{\prime}, \, \bm{q} ) .
\end{equation}
Now the scattering equation~\eqref{eq:LS-V} is expressed as an
equation of operators, as in Eq.~\eqref{eq:LS} in the nonrelativistic
case:
\begin{align}
\hat{\mathcal{T}} ( s )
= &
\hat{\mathcal{V}} ( s ) + \hat{\mathcal{V}} ( s ) 
\frac{1}{s - \hat{\mathcal{K}}} \hat{\mathcal{T}} ( s ) 
\notag \\
= &
\hat{\mathcal{V}} ( s ) + \hat{\mathcal{V}} ( s ) 
\frac{1}{s - \hat{\mathcal{K}} - \hat{\mathcal{V}} ( s )} 
\hat{\mathcal{V}} ( s ) .
\label{eq:LS-VII}
\end{align}
Actually, we can easily see that this operator equation becomes the
Lippmann--Schwinger equation~\eqref{eq:LS-V} by using the
normalization~\eqref{eq:norm_qco}.  In this sense, thanks to the
on-shell condition of the energy~\eqref{eq:q0_onshell},
Eqs.~\eqref{eq:LS-VII} and \eqref{eq:Psi_rel} become analogues to the
Lippmann--Schwinger equation and the \Schr equation in the
nonrelativistic case, respectively.

In this formulation, we can take the same strategy to calculate the
relativistic wave function as in the nonrelativistic case.  Near the
resonance pole, the scattering amplitude is dominated by the resonance
pole term in the expansion by the eigenstates of $\hat{\mathcal{K}} +
\hat{\mathcal{V}}$ as
\begin{equation}
\hat{\mathcal{T}} ( s ) \approx \sum _{M = - L}^{L} 
\hat{\mathcal{V}} ( s_{\rm pole} ) | \Psi _{L M} \rangle 
\frac{1}{s - s_{\rm pole}}
\langle \Psi _{L M}^{\ast} | \hat{\mathcal{V}} ( s_{\rm pole} ) ,
\label{eq:Tapprox-III}
\end{equation}
and hence the partial wave amplitude near the resonance pole position
is expressed as
\begin{equation}
T_{L, j k} ( s ; \, q^{\prime} , \, q ) = 
\frac{\gamma _{j} ( q^{\prime} ) \gamma _{k} ( q )}{s - s_{\rm pole}}
+ (\text{regular at } s = s_{\rm pole}) ,
\end{equation}
where we define the residue as 
\begin{equation}
\begin{split}
& {}_{\text{co}} \langle \bm{q}_{j} | 
\hat{\mathcal{V}} ( s_{\rm pole} ) | \Psi _{L M} \rangle
= \gamma _{j} ( q ) Y_{L M} ( \hat{q} ) , 
\\
& \langle \Psi _{L M}^{\ast} | \hat{\mathcal{V}} ( s_{\rm pole} ) 
| \bm{q}_{j} \rangle _{\text{co}}
= \gamma _{j} ( q ) Y_{L M}^{\ast} ( \hat{q} ) , 
\end{split}
\end{equation}
with 
\begin{equation}
\gamma _{j} ( q ) \equiv [ s_{\rm pole} - s_{j} ( q ) ] R_{j} ( q ) . 
\end{equation}
Then we can calculate the norm of the $j$th channel two-body state, as
the compositeness $X_{j}$, from the residue of the scattering
amplitude at the resonance pole:
\begin{align}
X_{j} = & 
\int \frac{d^{3} q}{( 2 \pi )^{3}} 
\frac{\sqrt{s_{j} ( q )}}
{2 \omega _{j} ( q ) \Omega _{j} ( q )}
\langle \Psi _{L M}^{\ast} | \bm{q}_{j} \rangle _{\text{co}} \; 
{}_{\text{co}} \langle \bm{q}_{j} | \Psi _{L M} \rangle 
\notag \\
= & \int \frac{d^{3} q}{( 2 \pi )^{3}} 
\frac{\sqrt{s_{j} ( q )}}
{2 \omega _{j} ( q ) \Omega _{j} ( q )}
\left [ \frac{\gamma _{j} ( q )}
{s_{\rm pole} - s_{j} ( q )} 
\right ]^{2} .
\label{eq:Xj_spinless}
\end{align}
This is the formula to evaluate the $j$th channel compositeness
$X_{j}$ from the residue of the partial wave amplitude $T_{L}$ at the
resonance pole for a relativistic scattering of two spinless
particles.

Finally let us consider a two-body relativistic scattering of
arbitrary spins.  The partial wave amplitude in this condition can be
specified by the orbital angular momentum $L$ and a certain index
$\alpha$ which represents quantum number of the scattering, such as
isospin.  The optical theorem for the partial wave amplitude is chosen
to be the same as that in Eq.~\eqref{eq:opt-rel}:
\begin{equation}
\text{Im} \, T_{\alpha L , \, j j}^{\text{on-shell}} ( w )
= - \sum _{k} \frac{q_{k} ( s )}{8 \pi w} 
\left | T_{\alpha L , j k}^{\text{on-shell}} ( w ) \right | ^{2} ,  
\end{equation}
where $w$ is the center-of-mass energy, $s \equiv w^{2}$, and the sum
runs over the open channels.  In general, the off-shell amplitude
$T_{\alpha L , \, j k}$ is a function of the center-of-mass energy $w$
and momenta $q^{\mu}$ and $q^{\prime \mu}$, but in order to relate the
scattering amplitude with the wave function clearly, we assume the
on-shell condition for the energy $q^{( \prime ) 0}$ so that it is a
function of the center-of-mass energy $w$ as in
Eq.~\eqref{eq:q0_onshell}.  Then, the partial wave amplitude can be
express near the resonance pole position as
\begin{equation}
T_{\alpha L, j k} ( w ; \, q^{\prime} , \, q ) = 
\frac{\gamma _{j}^{\prime} ( q^{\prime} ) \gamma _{k}^{\prime} ( q )}
     {w - w_{\rm pole}}
     + (\text{regular at } w = w_{\rm pole}) ,
\end{equation}
where $w_{\rm pole} \equiv \sqrt{s_{\rm pole}}$ is the pole position
in terms of the center-of-mass energy $w$.  Now we extend the
expression of the compositeness $X_{j}$ in the last line in
Eq.~\eqref{eq:Xj_spinless} to the scattering of arbitrary spin
particles.  Namely, in this study we define the compositeness for a
two-body system with arbitrary spin particles by using the residue of
the partial wave amplitude as
\begin{align}
X_{j} \equiv 2 w_{\rm pole}
\int \frac{d^{3} q}{( 2 \pi )^{3}} 
\frac{\sqrt{s_{j} ( q )}}
{2 \omega _{j} ( q ) \Omega _{j} ( q )}
\left [ \frac{\gamma _{j}^{\prime} ( q )}
{s_{\rm pole} - s_{j} ( q )} 
\right ]^{2} .
\label{eq:X_final}
\end{align}
This is the formula to evaluate the $j$th channel compositeness
$X_{j}$ from the resonance pole of the partial wave amplitude
$T_{\alpha L}$ for a relativistic scattering of arbitrary spin
particles.  We note that this formula of the compositeness is valid
even for baryonic resonances described with explicit Dirac gamma
matrices.  In the following we use this expression to evaluate the
compositeness of the $N^{\ast}$ resonances.

\subsection{Compositeness with separable interaction}
\label{sec:2C}

Up to now we have considered the nonrelativistic and relativistic
systems without specifying any explicit models for the interaction.
In the following we consider the interaction of the separable type,
i.e., the interaction $V ( w; \, q^{\prime} , \, q )$ which can be
factorized into the $q$ dependent part and $q^{\prime}$ dependent one.
The separable interaction is employed in the description of the
$N^{\ast}$ resonances in the chiral unitary approach in
Sec.~\ref{sec:3}.  We here concentrate on the scattering of the $\pi
N$ and other coupled channels in a relativistic framework, and hence
the partial wave amplitude is specified by isospin $I$, orbital
angular momentum $L$, and total angular momentum $J = L \pm 1/2$, as
$T_{I L}^{\pm}$.  In order to fix the interaction, we first note that
the radial wave function $R_{j} ( q )$ in $L$ wave behaves $\sim
q^{L}$ for the small $q$ region:
\begin{equation}
R_{j} ( q ) = \mathcal{O} ( q^{L} ) 
\quad 
\text{for small } q .
\end{equation}
Therefore, without loss of generality we can express the residue
of the partial wave amplitude $\gamma _{j}^{\prime} ( q )$ as
\begin{equation}
\gamma _{j}^{\prime} ( q ) = g_{j} q^{L} f_{j} ( q ) ,
\label{eq:gqf}
\end{equation}
where a constant $g_{j}$ is the coupling constant of the resonance to
the $j$th channel two-body state and a function $f_{j} ( q )$
satisfies $f_{j} ( 0 ) = 1$ and $f_{j} ( q ) \to 0$ for $q \to \infty$
so as to tame the ultraviolet divergence of the integrals.  Then it is
interesting that we can obtain the residue in Eq.~\eqref{eq:gqf}
exactly with the separable interaction of the following form:
\begin{equation}
V_{I L, j k}^{\pm} ( w; \, q^{\prime} , \, q )
= {V^{\prime}}_{I L, j k}^{\pm} ( w ) 
q^{\prime \, L} q^{L} 
f_{j} ( q^{\prime} ) f_{k} ( q ) ,
\end{equation}
where ${V^{\prime}}_{I L, j k}^{\pm}$ depends only on the
center-of-mass energy $w$.  This form of the interaction was proposed
in Refs.~\cite{Aceti:2012dd, Aceti:2014ala} so as to evaluate the
compositeness for higher partial wave states in a proper way.  With
this interaction, the full amplitude in $L$ wave can be obtained as
\begin{equation}
T_{I L, j k}^{\pm} ( w; \, q^{\prime} , \, q )
= {T^{\prime}}_{I L, j k}^{\pm} ( w ) q^{\prime \, L} q^{L} 
f_{j} ( q^{\prime} ) f_{k} ( q ) ,
\label{eq:separable_T}
\end{equation}
where ${T^{\prime}}_{I L, j k}^{\pm} ( w )$ is a solution of the
Lippmann--Schwinger equation in an algebraic form:
\begin{align}
  & {T^{\prime}}_{I L, j k}^{\pm} ( w )
  \notag \\
  & = {V^{\prime}}_{I L, j k}^{\pm} ( w ) 
  + \sum _{l} {V^{\prime}}_{I L, j l}^{\pm} ( w ) G_{L, l} ( w ) 
  {T^{\prime}}_{I L, l k}^{\pm} ( w ) .
\end{align}
In this expression, $G_{L, j}$ is the loop function of the two-body
state in $j$th channel, and in this study we take the following
expression
\begin{equation}
  G_{L, j} ( w ) \equiv \int \frac{d^{3} q}{( 2 \pi )^{3}}
\frac{\sqrt{s_{j} ( q )}}
{2 \omega _{j} ( q ) \Omega _{j} ( q )}
\frac{q^{2 L} [ f_{j} ( q ) ]^{2}}{s - s_{j} ( q )} ,
\end{equation}
in accordance with the discussion in Sec.~\ref{sec:2B}.  We note that,
in this construction, we have correct behavior of the on-shell
amplitude near the threshold:
\begin{equation}
T_{I L, j k}^{\pm} ( w ) ^{\text{on-shell}}
\propto q^{\prime \, L} q^{L} .
\end{equation}
An important point in this approach is that the scattering amplitude
${T^{\prime}}_{I L, j k}^{\pm}$, as well as the interaction
${V^{\prime}} _{I L, j k}^{\pm}$, depends only on the energy $w$.  Due
to this fact, the residue of ${T^{\prime}}_{I L, j k}^{\pm}$ at the
resonance pole position does not depend on the relative momentum $q$
and hence a constant:
\begin{equation}
{T^{\prime}}_{I L, j k}^{\pm} ( w ) =  
\frac{g_{j} g_{k}}{w - w_{\rm pole}} 
+ ( \text{regular at } w = w_{\rm pole} ) .
\end{equation}
One can easily confirm that the constant of the residue $g_{j}$
coincides with the prefactor in Eq.~\eqref{eq:gqf}.  Now, with the
full amplitude $T_{I L, j k}^{\pm}$ in Eq.~\eqref{eq:separable_T}, we
can straightforwardly calculate the compositeness~\eqref{eq:X_final}
in the present approach as
\begin{align}
X_{j} & = 2 w_{\rm pole} g_{j}^{2} \int \frac{d^{3} q}{(2 \pi )^{3}} 
\frac{\sqrt{s_{j} ( q )}}
{2 \omega _{j} ( q ) \Omega _{j} ( q )}
\left [ \frac{q^{L} f_{j} ( q )}
{s_{\rm pole} - s_{j} ( q )} \right ]^{2}
\notag \\
& = - g_{j}^{2} \left [ \frac{d G_{L, j}}{d w} \right ]_{w = w_{\rm pole}} ,
\label{eq:X}
\end{align}
where we have replaced the integral part in the middle of the equation
with the derivative of the loop function in the last.  Furthermore, in
the present approach we can express the elementariness $Z$ as
\begin{equation}
Z = - \sum _{j, k} g_{k} g_{j} 
\left [ G_{L, j} \frac{d {V^{\prime}}_{I L, j k}^{\pm} }{d w} G_{L, k} 
\right ]_{w = w_{\rm pole}} .
\end{equation}
Actually, with this expression we can show the normalization of the
total wave function for the resonance state:
\begin{align}
& \langle \Psi _{L M}^{\ast} | \Psi _{L M} \rangle 
= \sum _{j} X_{j} + Z 
\notag \\
& = - \sum _{j, k} g_{k} g_{j} 
\left [ \delta _{j k} \frac{d G_{L, j}}{d w} 
+ G_{L, j} \frac{d {V^{\prime}}_{I L, j k}^{\pm} }{d w} G_{L, k} 
\right ]_{w = w_{\rm pole}}
\notag \\
& = 1 ,
\end{align}
where the condition of the correct normalization as unity is
guaranteed by a generalized Ward identity proven in
Ref.~\cite{Sekihara:2010uz}.  The elementariness $Z$ measures the
contributions from missing channels which are effectively taken into
account in the two-body interaction in the practical model space,
including both one-body bare states and more than one-body scattering
states, on the assumption that the energy dependence of the
interaction originates from channels which do not appear as explicit
degrees of freedom.

Finally we note that in our numerical calculations we will employ the
dimensional regularization to calculate the integral of the loop
function, which is achieved by setting $f_{j} ( q ) = 1$ and modifying
the integration variable as $d^{4} k \to \mu _{\rm reg}^{4-d} d^{d}k$
with the regularization scale $\mu _{\rm reg}$.  The problems
concerned with the dimensional regularization will be discussed in the
next section.

\subsection{Interpretation of compositeness for resonances}
\label{sec:2D}

Before going to the numerical results of the compositeness for
$N^{\ast}$ resonances, we here give how to interpret and treat complex
values of the compositeness for resonance states, especially in the
relation to the probabilistic interpretation.  In this subsection we
consider a single channel problem for simplicity.  The extension to
the general coupled-channels case is given at the end.

As we have mentioned, the compositeness $X$ and elementariness $Z$ are
in general complex for resonance states, which is inevitable when the
correct normalization of the resonance wave function is required.
This fact indicates that we cannot interpret them as probabilities
since their values are not real and not bounded.  In this line,
several ways to make the compositeness real values have been proposed.
For instance, it has been suggested to use $1 - | Z
|$~\cite{Berggren}, $| X |$~\cite{Aceti:2012dd, Guo:2015daa},
$\text{Re}( X )$~\cite{Aceti:2014ala}, and $( 1 - | Z | + | X | ) /
2$~\cite{Kamiya:2015aea} as the ``probability'' of the compositeness.
All of them return to the same nonnegative value $1 - Z = X$ for a
stable bound state.  However, for a resonance state these values
except for the last one are not bounded in the range $[0, \, 1]$, so
the values $1 - | Z |$, $| X |$, and $\text{Re}( X )$ cannot be
treated as probability in a strict sense.

In contrast to these real values, in order to interpret the
compositeness and elementariness we propose to use simple but
reasonable values defined as
\begin{equation}
\tilde{X} \equiv \frac{ | X | }{1 + U} , \quad 
\tilde{Z} \equiv \frac{ | Z | }{1 + U} , 
\label{eq:Xtilde}
\end{equation}
with 
\begin{equation}
U \equiv | X | + | Z | - 1 . 
\end{equation}
Obviously, both $\tilde{X}$ and $\tilde{Z}$ are real, bounded in the
range $[0, \, 1]$, and automatically satisfy the sum rule:
\begin{equation}
\tilde{X} + \tilde{Z} = 1 .
\end{equation}
We then require that we can interpret $\tilde{X}$ and $\tilde{Z}$ from
the complex compositeness and elementariness as the ``probability'' if
and only if $U$ is much smaller than unity, $U \ll 1$.  This is
essentially an expression of the condition pointed out in
Ref.~\cite{Sekihara:2014kya}, in which they proposed that reasonable
interpretation can be obtained if $| \text{Im}(Z) |$, $| \text{Im}(X)
| \ll 1$ and $0 \lesssim \text{Re}(Z)$, $\text{Re}(X) \lesssim 1$,
where $X$ and $Z$ have similarity with those of the stable bound
states.  In other words, we can have a resonance wave function which
is similar to the wave function of the stable bound state.  Here we
note that, when $U \ll 1$ is satisfied, $\tilde{X}$ and $\tilde{Z}$ in
Eq.~\eqref{eq:Xtilde} take very similar values to the quantities
proposed in Ref.~\cite{Kamiya:2015aea}:
\begin{equation}
\tilde{X}_{\rm KH} \equiv \frac{1 - | Z | + | X |}{2} , 
\quad 
\tilde{Z}_{\rm KH} \equiv \frac{1 - | X | + | Z |}{2} . 
\end{equation}
Actually, a straightforward calculation provides
\begin{align}
\tilde{X} - \tilde{X}_{\rm KH} 
= & \frac{|X|}{1 + U} - \frac{1 - | Z | +  | X |}{2} 
= \frac{ ( | Z | - | X | ) U}{2 ( 1 + U )} ,
\end{align}
which should be much smaller than unity for $U \ll 1$.  A similar
result for $\tilde{Z} - \tilde{Z}_{\rm KH}$ is obtained by exchanging
$| X |$ with $| Z |$.  We also mention that, with the condition $U \ll
1$, the quantities $1 - | Z |$, $| X |$, and $\text{Re}( X )$ will
take similar values to $\tilde{X}$.

Finally, an important property is that we can straightforwardly extend
$\tilde{X}$ and $\tilde{Z}$ to the general coupled-channel case.  This
can be done as
\begin{equation}
\tilde{X}_{j} \equiv \frac{ | X_{j} | }{1 + U} , \quad 
\tilde{Z} \equiv \frac{ | Z | }{1 + U} , 
\end{equation}
with 
\begin{equation}
U \equiv \sum _{j} | X_{j} | + | Z | - 1 . 
\end{equation}
Again $\tilde{X}_{j}$ and $\tilde{Z}$ are real, bounded in the range
$[0, \, 1]$, and automatically satisfy the sum rule:
\begin{equation}
\sum _{j} \tilde{X}_{j} + \tilde{Z} = 1 .
\end{equation}
In the following we will use these real values as well as the original
compositeness $X_{j}$ and elementariness $Z$ when we discuss the
internal structure of $N^{\ast}$ resonances.

\section{Numerical results}
\label{sec:3}

Let us now consider the $\Delta (1232)$, $N (1535)$, and $N (1650)$
resonances and evaluate their meson--baryon compositeness from the
residue of the scattering amplitude at the resonance pole position by
using the formula developed in Sec.~\ref{sec:2C}.  We employ the
chiral unitary approach to calculate the scattering amplitude.  The
chiral unitary approach is most successful in description of the
$\Lambda (1405)$ resonance~\cite{Kaiser:1995eg, Oset:1997it,
  Oller:2000fj, Lutz:2001yb, GarciaRecio:2002td, Jido:2003cb,
  Hyodo:2011ur}, and is applied to the $\pi N$ scattering and several
$N ^{\ast}$ resonances as well~\cite{Kaiser:1995cy, Meissner:1999vr,
  Nieves:2001wt, Inoue:2001ip, Hyodo:2002pk, Hyodo:2003qa,
  Bruns:2010sv, Alarcon:2012kn, Khemchandani:2013nma, Garzon:2014ida}.
In this study, the interaction kernel is taken from chiral
perturbation theory up to the next-to-leading order, and we construct
separable interactions to evaluate the $\pi N$ scattering amplitude.
The loop function is evaluated with the dimensional regularization.
The model parameters are fitted so that the partial wave amplitudes
reproduce the solution of the partial wave analysis for the $\pi N$
scattering amplitude obtained in Ref.~\cite{Workman:2012hx}, to which
we refer as WI~08.  Throughout the numerical calculations we use
isospin symmetric masses for mesons and baryons.

\subsection{The $\bm{\Delta (1232)}$ resonance}
\label{sec:3A}

First we consider the $\Delta (1232)$ resonance and calculate its $\pi
N$ compositeness.  In this study we construct the $\pi N$
single-channel scattering amplitude in $s$ and $p$ waves by using the
unitarization of the chiral interaction up to the next-to-leading
order plus the $s$- and $u$-channel $\Delta (1232)$ exchanges.  In the
analysis we also consider the $\pi N$ compositeness for the ground
state nucleon $N (940)$.

Since both $N (940)$ and $\Delta (1232)$ exist in $p$-wave $\pi N$
state with the orbital angular momentum $L = 1$, we have to employ the
loop function with $L = 1$.  As we will see, there is ambiguity in
calculating the $\pi N$ compositeness with the $L = 1$ loop function
evaluated with the dimensional regularization.  We discuss this
ambiguity as well.

\subsubsection{Scattering amplitude}

Let us consider the $\pi N$ scattering amplitude in isospin $I$,
orbital angular momentum $L$, and total angular momentum $J = L \pm
1/2$, which is denoted by $T_{I L}^{\pm} ( w; \, \bm{q}^{\prime} , \,
\bm{q} )$ as a function of the center-of-mass energy $w$ and relative
momentum in the initial (final) state $\bm{q}^{( \prime )}$.  An
important property of the scattering amplitude $T_{I L}^{\pm}$ is
that, for each isospin and angular momentum, it should satisfy the
optical theorem from the unitarity of the scattering matrix.  Namely,
below the inelastic threshold, the on-shell amplitude should satisfy
\begin{align}
\text{Im} \, T_{I L}^{\pm} ( w )^{\text{on-shell}}
= & - \frac{\rho _{\pi N} ( s )}{2} 
\left | T_{I L}^{\pm} ( w )^{\text{on-shell}} \right | ^{2}
\notag \\
& \times \theta ( w - m_{\pi} - M_{N} ) ,  
\label{eq:optical}
\end{align}
where $s \equiv w^{2}$, $\theta (x)$ is the Heaviside step function,
$m_{\pi}$ and $M_{N}$ are the pion and nucleon masses, respectively,
and $\rho _{\pi N} ( s )$ is the phase space defined as
\begin{equation}
  \rho _{\pi N} ( s ) \equiv
  \frac{q_{\pi N} ( s )}{4 \pi \sqrt{s}},
  \quad 
  q_{\pi N} ( s ) \equiv
  \frac{\lambda ^{1/2} ( s , \, m_{\pi}^{2} , \, M_{N}^{2} )}
       {2 \sqrt{s}} ,
\end{equation}
with the \Kaellen function $\lambda (x, \, y, \, z) = x^{2} + y^{2} +
z^{2} - 2 x y - 2 y z - 2 z x$.  The optical
theorem~\eqref{eq:optical} can be equivalently rewritten in the
following form:
\begin{equation}
\text{Im} \left [ T_{I L}^{\pm} ( w )^{\text{on-shell}} \right ] ^{-1} 
= \frac{\rho _{\pi N} (s)}{2} \theta ( w - m_{\pi} - M_{N} ) .
\label{eq:optical-II}
\end{equation}

\begin{figure}
  \centering
  \PsfigII{0.22}{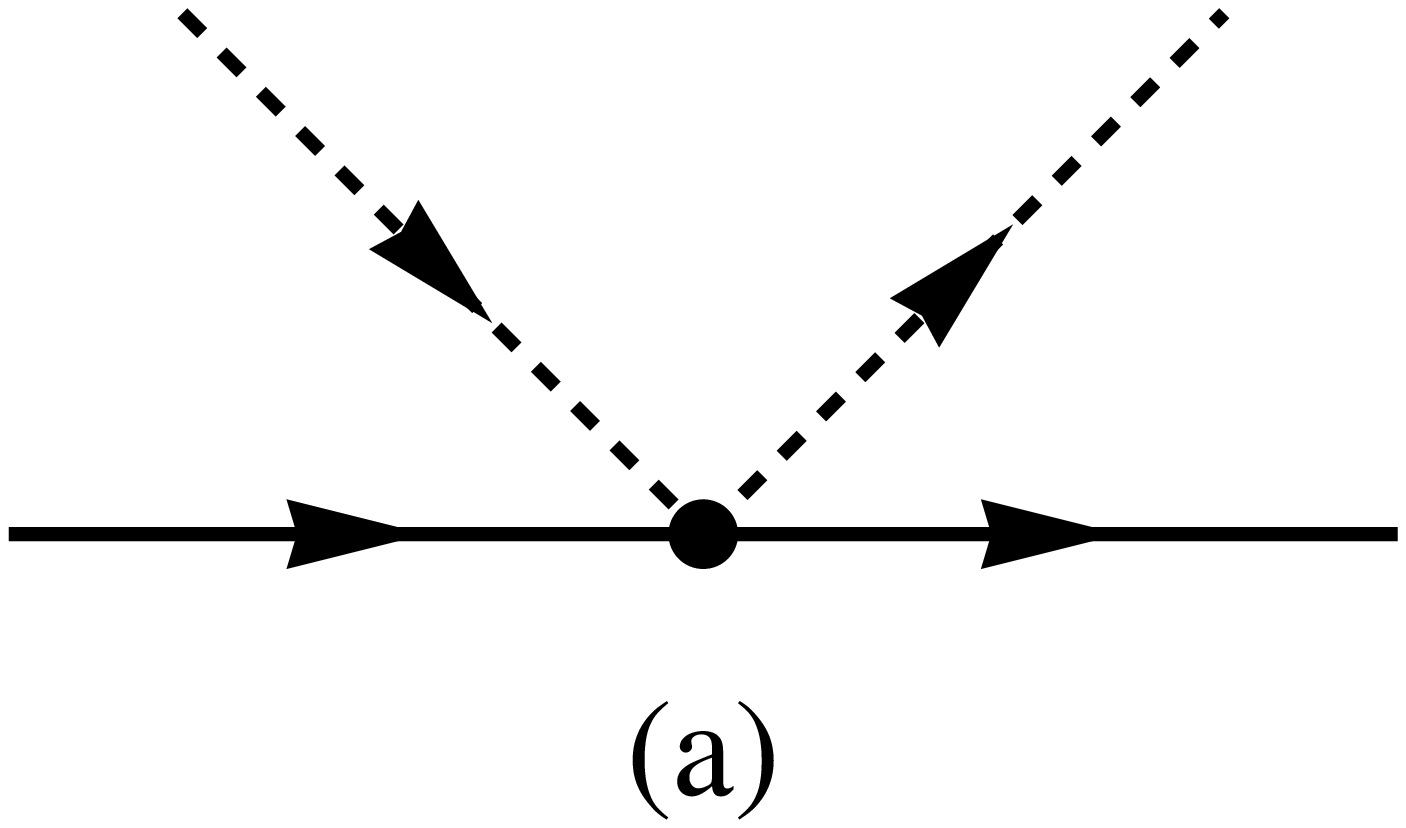}~\hspace{20pt}~\PsfigII{0.22}{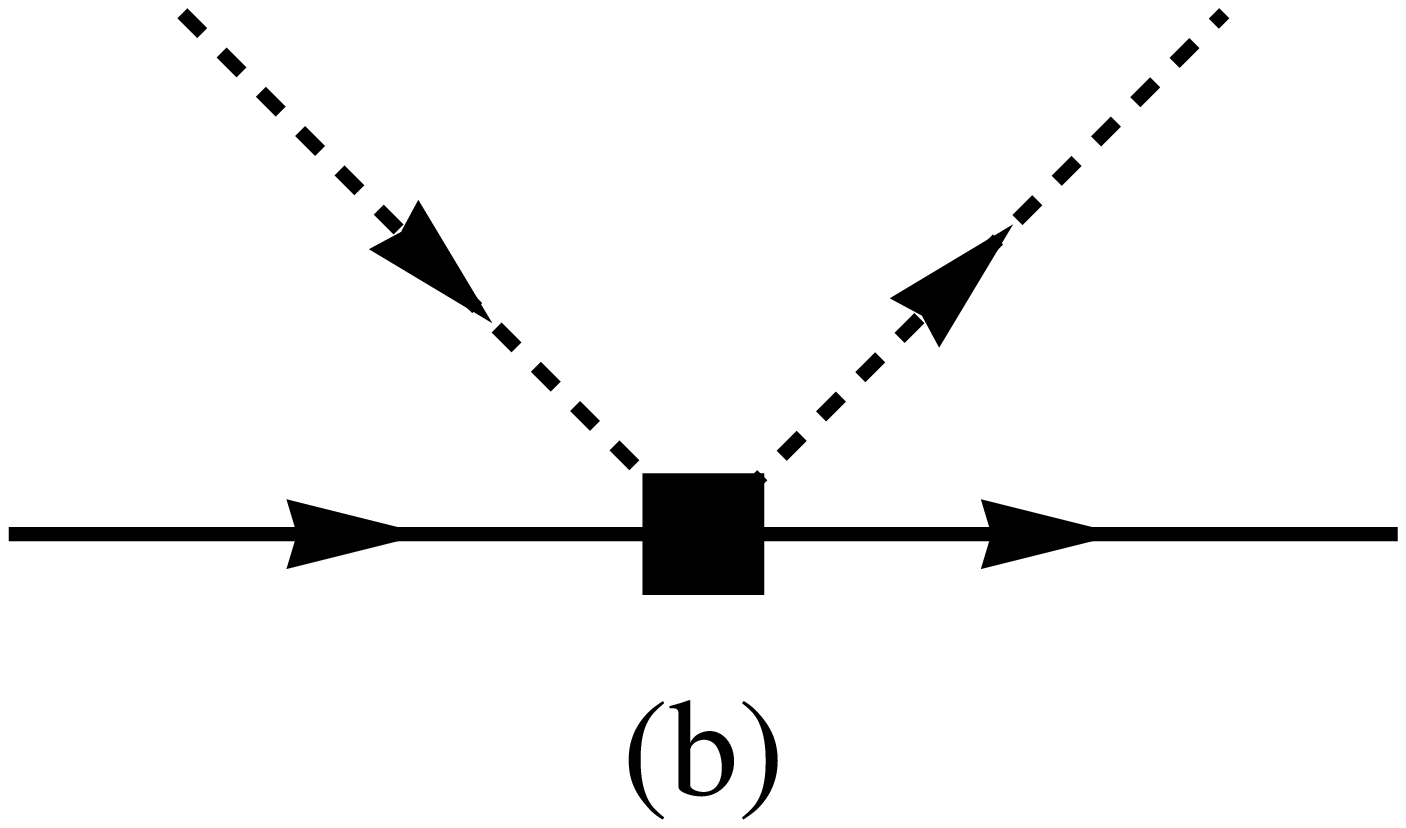} \\
  \PsfigII{0.22}{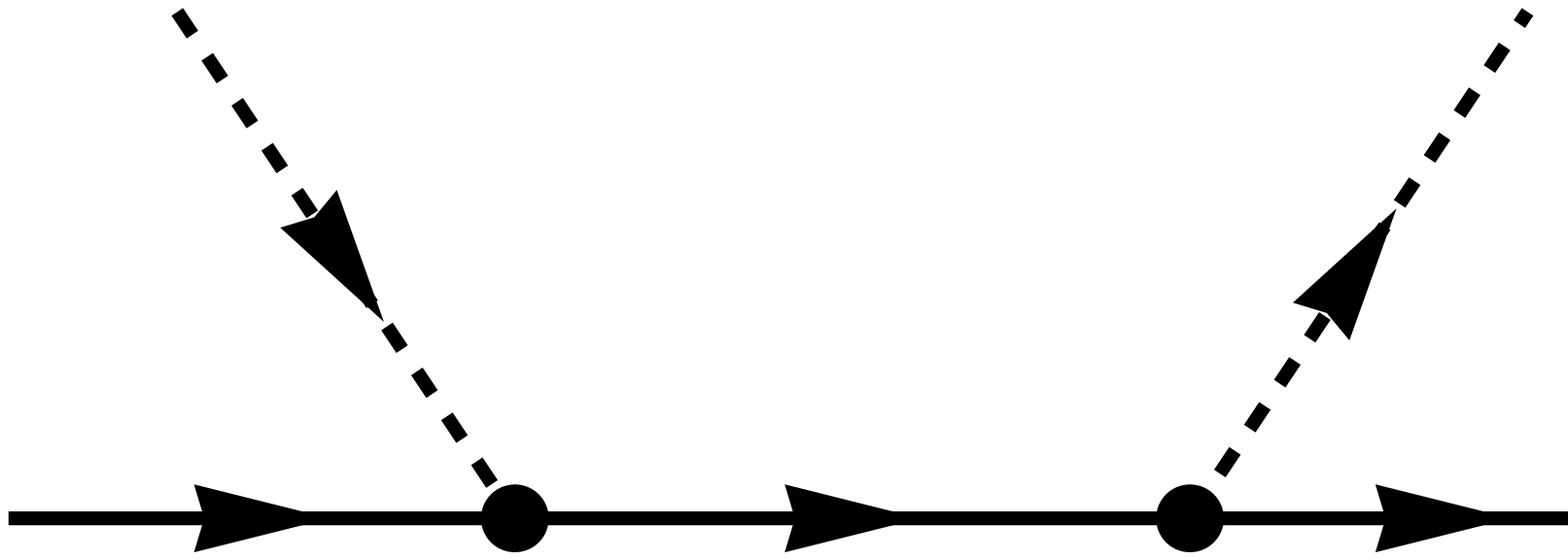} \\
  \PsfigII{0.22}{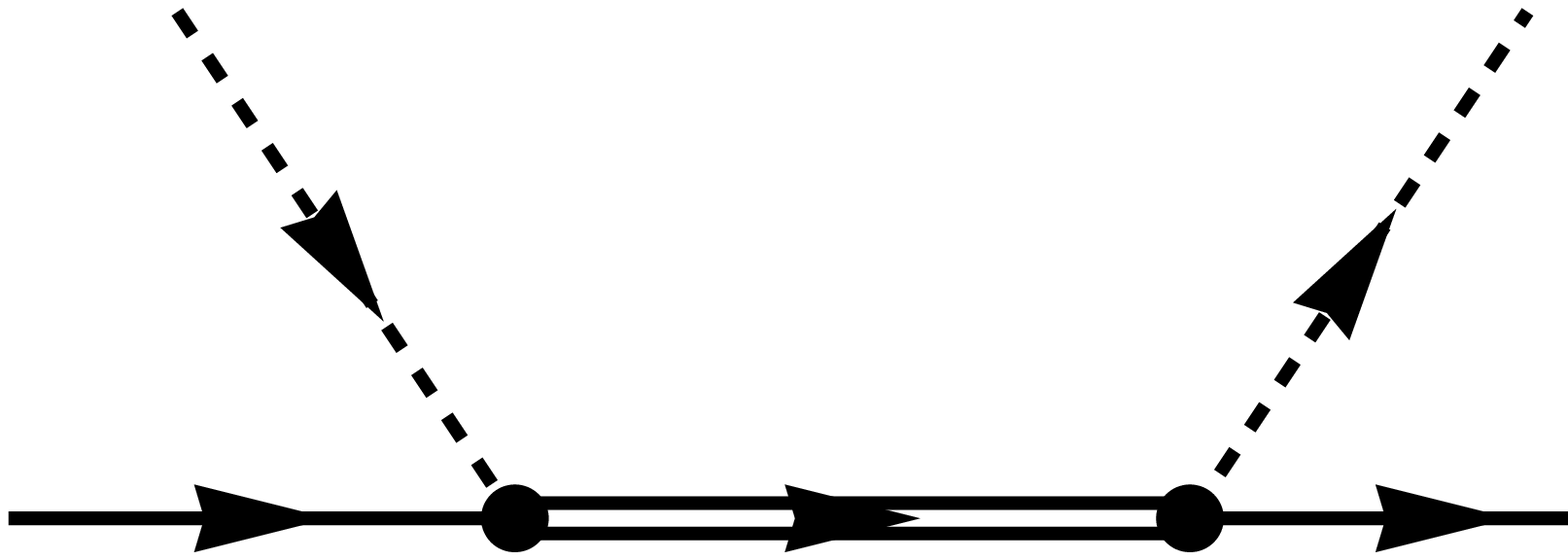}
  \caption{Feynman diagrams for the interaction kernel: (a)
    Weinberg--Tomozawa term, (b) next-to-leading order term, (c) $s$-
    and $u$-channel $N (940)$ exchange terms, and (d) $s$- and
    $u$-channel $\Delta (1232)$ exchange terms.  The solid, dashed,
    and double lines represent baryons, mesons, and $\Delta (1232)$,
    respectively.  The dots and square represent the $\mathcal{O}
    (p^{1})$ and $\mathcal{O} (p^{2})$ vertices from chiral
    perturbation theory, respectively.}
  \label{fig:V}
\end{figure}

The chiral unitary approach is a model to construct the scattering
amplitude which satisfies the optical theorem~\eqref{eq:optical} with
the interaction taken from chiral perturbation theory.  In order to
formulate the chiral unitary approach, we first fix the interaction
kernel for the scattering equation.  In this study we employ chiral
perturbation theory up to $\mathcal{O} (p^{2})$ for the $\pi N$
interaction kernel $V_{I L}^{\pm}$.  The interaction kernel consists
of the Weinberg--Tomozawa term $V_{\rm WT}$, $s$- and $u$-channel
$N(940)$ exchanges $V_{s+u}$, next-to-leading order contact term
$V_{2}$, and $s$- and $u$-channel $\Delta (1232)$ exchanges
$V_{\Delta}$ (see Fig.~\ref{fig:V}).  They are projected to the
partial wave components as
\begin{equation}
  V_{I L}^{\pm} ( w; \, | \bm{q}^{\prime} | , \, | \bm{q} | )
  = [ V_{\rm WT} + V_{s+u} + V_{2} + V_{\Delta} ]_{I L}^{\pm} .
  \label{eq:Vproj}
\end{equation}
The explicit expression of each term is given in Appendix~\ref{app:2}.
The interaction kernel has six model parameters altogether: the
low-energy constants $c_{1}$, $c_{2}$, $c_{3}$, and $c_{4}$, the bare
$\Delta$ mass $M_{\Delta}$, and the $\pi N \Delta$ bare coupling
constant $g_{\pi N \Delta}$.  Then, according to the discussion in
Sec.~\ref{sec:2C}, we factorize the relative momenta of the order of
the orbital angular momentum $| \bm{q}^{\prime} |^{L} | \bm{q} |^{L}$,
which is essential to evaluate the compositeness for higher partial
wave states~\cite{Aceti:2012dd, Aceti:2014ala}, as
\begin{equation}
V_{I L}^{\pm} ( w; \, | \bm{q}^{\prime} | , \, | \bm{q} | ) 
= | \bm{q}^{\prime} |^{L} | \bm{q} |^{L} 
{V^{\prime}}_{I L}^{\pm} ( w ) ,
\label{eq:separable_Delta}
\end{equation}
where we have applied the on-shell condition to ${V^{\prime}}_{I
  L}^{\pm}$ so that it depends only on the center-of-mass energy $w$,
by replacing the pion momentum $q^{\mu}$ in ${V^{\prime}}_{I L}^{\pm}$
with the corresponding on-shell values:
\begin{equation}
  q^{0} \to \omega ( w ) \equiv \frac{s + m_{\pi}^{2} - M_{N}^{2}}{2 w},
  \quad
  | \bm{q} | \to q_{\pi N} ( s ) .
\end{equation}
With this interaction kernel, the full scattering amplitude can be
obtained as
\begin{equation}
T_{I L}^{\pm} ( w; \, | \bm{q}^{\prime} | , \, | \bm{q} | ) 
= | \bm{q}^{\prime} |^{L} | \bm{q} |^{L} 
{T^{\prime}}_{I L}^{\pm} ( w ) .
\end{equation}
where ${T^{\prime}}_{I L}^{\pm} ( w )$ is a solution of the
Lippmann--Schwinger equation in an algebraic form but without the
factor of the external momenta $| \bm{q}^{\prime} |^{L} | \bm{q}
|^{L}$:
\begin{equation}
{T^{\prime}}_{I L}^{\pm} ( w ) 
= {V^{\prime}}_{I L}^{\pm} + {V^{\prime}}_{I L}^{\pm} G_{L} {T^{\prime}}_{I L}^{\pm} 
= \frac{1}{1/{V^{\prime}}_{I L}^{\pm} ( w ) - G_{L} ( w )} . 
\end{equation}
Here, $G_{L}$ is the loop function containing the contribution from
the internal momentum $| \bm{q} |^{2 L}$, and in this study we take
the following expression:
\begin{equation}
G_{L} ( w ) = i \int \frac{d^{4} q}{( 2 \pi )^{4}}
\frac{| \bm{q} |^{2 L}}{( q^{2} - m_{\pi}^{2} ) [ (P - q)^{2} - M_{N}^{2} ]} ,
\label{eq:GL}
\end{equation}
with $P^{\mu} = ( w, \, \bm{0} )$.

Next let us focus on the loop function $G_{L}$.  In this study we
evaluate the loop function with the subtraction scheme, rather than a
cutoff, and the dimensional regularization.  Since the loop function
contains the internal momentum $| \bm{q} |^{2 L}$, the integral in
Eq.~\eqref{eq:GL} diverges logarithmically for $L = 0$ and it becomes
worse for $L > 0$.  Therefore, in order to make the integral finite,
we need to subtract the divergences $L + 1$ times in the subtraction
scheme (see Appendix~\ref{app:3} for the details).  For instance, the
$L = 0$ loop function is evaluated with a subtraction constant $a$ as
\begin{align}
& G_{L = 0} (w ; \, a)  \equiv 
\frac{1}{16 \pi ^{2}} \left [ 
a + \frac{s + m_{\pi}^{2} - M_{N}^{2}}{2 s} 
\ln \left ( \frac{m_{\pi}^{2}}{M_{N}^{2}} \right )
\right .
\notag \\ 
& - \frac{\lambda ^{1/2} (s, \, m_{\pi}^{2}, \, M_{N}^{2})}{s} 
\text{artanh} \left . \!
\left ( \frac{\lambda ^{1/2} (s, \, m_{\pi}^{2}, \, M_{N}^{2})}
{m_{\pi}^{2} + M_{N}^{2} - s} \right ) 
\right ] ,
\label{eq:GL0_explicit}
\end{align}
where the regularization scale is fixed as $\mu _{\rm reg} = M_{N}$,
as in Appendix~\ref{app:3}.  On the other hand, when we calculate the
$L = 1$ loop function for $\Delta (1232)$ in the $P_{33}$ amplitude
and for $N (940)$ in $P_{11}$, we need two subtraction constants.  We
now eliminate one of the two subtraction constants by requiring that
the nucleon pole does not shift in the unitarization of the $\pi N$
scattering amplitude in $P_{11}$, which constrains the $L = 1$ loop
function as
\begin{equation}
  G_{L = 1} ( M_{N} ) = 0 .
  \label{eq:GL1_condition}
\end{equation}
Physically, this means that we do not perform the renormalization of
the nucleon mass, but the wave function renormalization of the nucleon
is allowed to take place since $d G_{L = 1} / d w ( M_{N} )$ may not
be zero.  As derived in Appendix~\ref{app:3}, the
condition~\eqref{eq:GL1_condition} brings the loop function in the
following expression:
\begin{equation}
G_{L = 1} ( w ) = G_{\pi N, \, L = 1} ( w ; \, \tilde{A} ) ,
\end{equation}
\begin{align}
& G_{\pi N, \, L = 1} ( w ; \, \tilde{A} ) = 
\frac{s - M_{N}^{2}}{4} \tilde{A}
+ \frac{s G_{\pi N} ( w )}{4} 
\notag \\
& 
- \frac{m_{\pi}^{2} + M_{N}^{2}}{2} G_{\pi N} ( w )
\notag \\
& + \frac{( m_{\pi}^{2} - M_{N}^{2} )^{2}}{4} 
\left [ \frac{G_{\pi N} ( w ) - G_{\pi N} ( 0 )}{s} 
+ \frac{G_{\pi N} ( 0 )}{M_{N}^{2}} \right ] .
\label{eq:GL_dim}
\end{align}
Here, $\tilde{A}$ is the remaining subtraction constant, which becomes
a model parameter, and $G_{\pi N} ( w )$ is the $L = 0$ loop function
with the condition $G_{\pi N} ( M_{N} ) = 0$:
\begin{align}
& G_{\pi N} ( w ) = G_{L = 0} ( w ; \, 0 ) - G_{L = 0} ( M_{N} ; \, 0 ) .
\end{align}
One can easily check that the loop function $G_{\pi N, \, L = 1}$
satisfies $G_{\pi N, \, L = 1} ( M_{N} ) = 0$.  Besides, we note that
the $S_{11}$ and $S_{31}$ amplitudes are not important in the study on
$\Delta (1232)$.  Therefore, to calculate the $S_{11}$ and $S_{31}$
amplitudes we also require the $L = 0$ loop function to be zero at $w
= M_{N}$, for simplicity:
\begin{equation}
G_{L = 0} ( w ) = G_{\pi N} ( w ) .
\end{equation}
This condition is achieved also by the natural renormalization
scheme~\cite{Hyodo:2008xr}, which can exclude explicit pole
contributions from the loop functions.

\def\arraystretch{1.2}

\begin{table}[!t]
  \caption{Fitted parameters for the $\pi N$ amplitudes $S_{11}$,
    $S_{31}$, $P_{11}$, $P_{31}$, $P_{13}$, and $P_{33}$.  We also
    show the $\chi ^{2}$ value divided by the number of degrees of
    freedom, $\chi ^{2} / N_{\rm d.o.f.}$. }
  \label{tab:1}
  \begin{ruledtabular}
    \begin{tabular*}{8.6cm}{@{\extracolsep{\fill}}lcc}
      & Naive & Constrained \\
      \hline
      $c_{1}$ [GeV${}^{-1}$] & $-0.111\phantom{-}$ & 
      $-0.047\phantom{-}$
      \\
      $c_{2}$ [GeV${}^{-1}$] & $0.725$ & 
      $0.810$
      \\
      $c_{3}$ [GeV${}^{-1}$] & $-1.797\phantom{-}$ & 
      $-1.784 \phantom{-}$ 
      \\
      $c_{4}$ [GeV${}^{-1}$] & $0.089$ & 
      $0.512$
      \\
      $g_{\pi N \Delta}$ & $1.808$ & $1.507$ \\
      $M_{\Delta}$ [MeV] & $1296.0$ & $1320.6$ \\
      $\tilde{A}$ & $-3.61 \times 10^{-3}$ & 
      $- 4.82\times 10^{-3}$ \\
      \\
      $\chi ^{2} / N_{\rm d.o.f.}$ & 
      $486.3 / 809$
      & 
      $1239.9/809$
      \\
    \end{tabular*}
  \end{ruledtabular}
\end{table}

\begin{figure}[!t]
  \centering
  \Psfig{8.6cm}{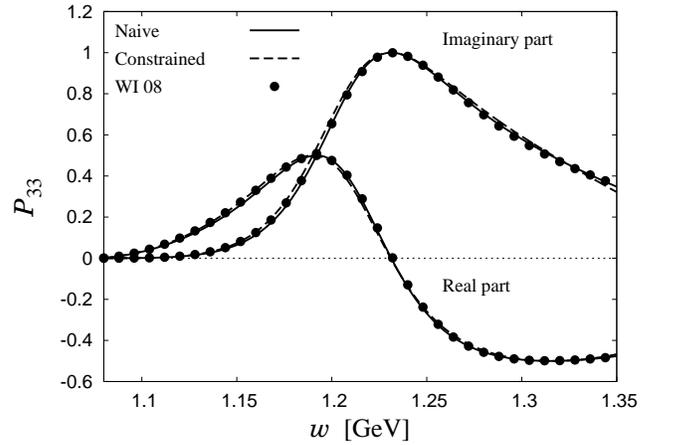}
  \caption{Scattering amplitude $P_{33}$ with parameter sets Naive and
    Constrained fitted to the WI~08 solution~\cite{Workman:2012hx}.
    Two theoretical curves are very similar.  The number of the
    plotted data points is $1/2$ of the total in the fits for a better
    visualization.}
  \label{fig:P33}
\end{figure}

Now we have the formulation to calculate the scattering amplitude for
$\Delta (1232)$ in the chiral unitary approach.  In the present
formulation, we have seven model parameters.  They are fixed so as to
reproduce the solution of the $\pi N$ partial wave analysis
WI~08~\cite{Workman:2012hx}.  In the fitting procedure, we introduce a
normalized scattering amplitude
\begin{equation}
L_{2 I \, 2 J} ( w ) 
= - \frac{\rho _{\pi N} (s) q_{\pi N} (s)^{2 L}}{2} 
{T^{\prime}}_{I L}^{\pm} (w)^{\text{on-shell}} ,
\end{equation}
which satisfies the following optical theorem:
\begin{equation}
\text{Im} \, L_{2 I \, 2 J} ( w ) = | L_{2 I \, 2 J} ( w ) |^{2}
\theta ( w - m_{\pi} - M_{N} ) ,
\label{eq:optL}
\end{equation}
below the inelastic threshold for the $\pi N$ state.  We fit six $\pi
N$ amplitudes $S_{11} (w)$, $S_{31} (w)$, $P_{11} (w)$, $P_{31} (w)$,
$P_{13} (w)$, and $P_{33} (w)$ to the WI~08 solution up to $1.35 \gev$
in intervals of $4 \mev$, in which only $\Delta (1232)$ appears as the
$N^{\ast}$ resonance.\footnote{The present energy range is even below
  the first excitation in $P_{11}$, i.e., the Roper resonance.
  Nevertheless, we can in principle calculate the compositeness for
  the Roper resonance by introducing scattering states of higher
  thresholds relevant to the Roper resonance and by fitting higher
  energy regions as well.}  We note that the WI~08 solution does not
provide errors for the scattering amplitude.  For the calculation of
the $\chi ^{2}$ value, in this study we introduce a common error
$0.01$ both for the real and imaginary parts of the scattering
amplitude in every quantum number.  From the best fit to the WI~08
solution, we obtain the model parameters listed in the second column
of Table~\ref{tab:1}, to which we refer as the ``Naive'' parameters.
We mention that the low-energy constants found in this fitting are in
general not identical to the ones from tree-level chiral perturbation
theory, since we have fit them to the scattering amplitude including
the $\Delta (1232)$ resonance region rather than fit them to the
masses of baryons nor to low-energy phenomena around the $\pi N$
threshold.  We also show the $P_{33}$ amplitude in the theoretical
calculation and the WI~08 solution in Fig.~\ref{fig:P33}, which shows
a good reproduction of the $P_{33}$ amplitude by the parameter set
Naive.

\subsubsection{Compositeness}

Now that we have determined the scattering amplitude, let us evaluate
the $\pi N$ compositeness for $\Delta (1232)$ and $N (940)$ from their
pole positions and residues.  In the present formulation, the
scattering amplitude has the resonance pole in the following
expression:
\begin{equation}
{T^{\prime}}_{I L}^{\pm} ( w ) = \frac{g^{2}}{w - w_{\rm pole}} 
+ (\text{regular at } w = w_{\rm pole}) ,
\end{equation}
where $g$ is the coupling constant of the resonance to the $\pi N$
state and $w_{\rm pole}$ is the pole position in the complex $w$
plane.  We note that $g$ and $w_{\rm pole}$ contains information on
the structure of the resonance, and this is formulated in terms of the
compositeness as developed in Sec.~\ref{sec:2C}:
\begin{equation}
X_{\pi N} = - g^{2} \frac{d G_{L}}{d w} ( w = w_{\rm pole} ) ,
\end{equation}
which measures the amount of the two-body composite fraction inside
the resonance.  In addition, we can calculate the elementariness as
well:
\begin{equation}
  Z = - g^{2} 
  \left [ 
    G_{L}^{2} \frac{d {V^{\prime}}_{I L}^{\pm}}{d w} 
  \right ]_{w = w_{\rm pole}} .
\label{eq:Z-gp}
\end{equation}
The elementariness $Z$ measures the contributions from missing
channels which are effectively taken into account in the $\pi N$
interaction in the practical model space, on the assumption that the
energy dependence of the interaction originates from channels which do
not appear as explicit degrees of freedom.  It is important that we
have the normalization of the total wave function as
\begin{equation}
X_{\pi N} + Z = 1 .
\end{equation}
However, in general, both the compositeness $X_{\pi N}$ and the
elementariness $Z$ are complex for the resonance states, which are
difficult to interpret.  Therefore, we introduce quantities which are
real, bounded in the range $[0, \, 1]$, and automatically satisfy the
sum rule:
\begin{equation}
\tilde{X}_{\pi N} \equiv \frac{ | X_{\pi N} | }{1 + U} , \quad 
\tilde{Z} \equiv \frac{ | Z | }{1 + U} , 
\end{equation}
with 
\begin{equation}
U \equiv | X_{\pi N} | + | Z | - 1 . 
\end{equation}
Obviously, we have the sum rule for $\tilde{X}_{\pi N}$ and
$\tilde{Z}$:
\begin{equation}
\tilde{X}_{\pi N} + \tilde{Z} = 1 .
\end{equation}
We can interpret $\tilde{X}_{\pi N}$ and $\tilde{Z}$ from the complex
compositeness and elementariness as the ``probability'' if and only if
$U$ is much smaller than unity, $U \ll 1$.

\begin{table}[!t]
  \caption{Properties of $\Delta (1232)$ and $N(940)$.  We do not
    calculate $U$, $\tilde{X}_{\pi N}$, and $\tilde{Z}$ for $N (940)$
    since it is a stable state.}
  \label{tab:2}
  \begin{ruledtabular}
    \begin{tabular*}{8.6cm}{@{\extracolsep{\fill}}lcccc}
      & \multicolumn{2}{c}{Naive}
      & \multicolumn{2}{c}{Constrained} \\
      & $\Delta (1232)$ & $N (940)$ 
      & $\Delta (1232)$ & $N (940)$
      \\
      \hline
      $w_{\rm pole}$ [MeV]
      & $1209.8 - 47.6 i$ & $938.9$
      & $1206.9 - 49.6 i$ & $938.9$ 
      \\
      $g$ [MeV${}^{-1/2}$]
      & $0.383 - 0.053 i$ & $0.560$
      & $0.395 - 0.061 i$ & $0.516$
      \\
      $X_{\pi N}$
      & $0.69 + 0.39 i$ & $-0.18\phantom{-}$
      & $0.87 + 0.35 i$ & $0.00$ 
      \\
      $Z$
      & $0.31 - 0.39 i$ & $1.18$ 
      & $0.13 - 0.35 i$ & $1.00$
      \\
      $U$
      & $0.30$ & ---
      & $0.31$ & ---
      \\
      $\tilde{X}_{\pi N}$
      & $0.61$ & ---
      & $0.71$ & ---
      \\
      $\tilde{Z}$
      & $0.39$ & ---
      & $0.29$ & --- 
      \\
    \end{tabular*}
  \end{ruledtabular}
\end{table}

Now we calculate the pole positions, coupling constants,
compositeness, and elementariness in the parameter set Naive, and list
them in the second and third columns of Table~\ref{tab:2}.  First, the
$\Delta (1232)$ pole position in the parameter set Naive is very
similar to that reported by Particle Data Group: $w_{\rm pole} = (1210
\pm 1) - (50 \pm 1) i \mev$~\cite{Olive:1900zz}.  The $\pi N$
compositeness is evaluated as $X_{\pi N} = 0.69 + 0.39 i$, which
implies that the $\Delta (1232)$ resonance contains a significant $\pi
N$ component.  Thus, our result in the refined model reconfirms the
calculation in Ref.~\cite{Aceti:2014ala}.  We note that the imaginary
part of the compositeness is nonnegligible as well, but the value of
$U = 0.30$ is less than one third.  This implies that we may interpret
$\tilde{X}_{\pi N}$ and $\tilde{Z}$ as the ``probability'' to find the
$\pi N$ composite and missing-channel contributions, respectively.
From the values of $\tilde{X}_{\pi N}$ and $\tilde{Z}$, we may
conclude that the $\Delta (1232)$ resonance in the present refined
model contains a significant $\pi N$ component.

\begin{figure}[!t]
  \centering
  \Psfig{8.6cm}{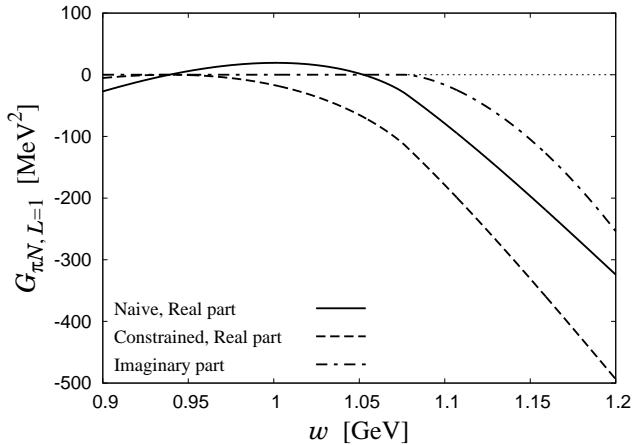}
  \caption{Loop function $G_{\pi N, \, L = 1}$ around the energy from
    the nucleon mass $M_{N}$ to the $\pi N$ threshold.  The imaginary
    part of the loop function takes the same value both in the
    parameter sets Naive and Constrained.}
  \label{fig:Gloop}
\end{figure}

On the other hand, for $N (940)$, the wave function renormalization
takes place due to $d G_{L = 1} / d w ( M_{N} ) \ne 0$, and its $\pi N$
compositeness becomes finite in the parameter set Naive.  However, its
value is real but negative.  This result is unphysical, since we
cannot interpret it as a probability although $N (940)$ is a stable
state.  The origin of the negative compositeness is the fact that the
derivative of the $L = 1$ loop function, $d G_{L=1} / d w$, becomes
positive at the nucleon pole $w = M_{N}$, which can be seen in
Fig.~\ref{fig:Gloop} (solid line).  As a result, the compositeness
becomes negative even when the coupling constant is real: $g = 0.560
\mev ^{-1/2}$.

Physically the derivative of the loop function, $d G_{L} / d w$,
should not be positive below the $\pi N$ threshold.  We can see this
by looking into the expression of the loop function $G_{L}$
(see Appendix~\ref{app:3}):
\begin{equation}
  G_{L} ( w ) \equiv 
  - \int _{s_{\rm th}}^{\infty} \frac{d s^{\prime}}{2 \pi} 
  \frac{\rho _{\pi N} (s^{\prime}) q_{\pi N} ( s^{\prime} )^{2 L}}
       {s^{\prime} - s - i 0} .
\end{equation}
with $s_{\rm th} \equiv ( m_{\pi} + M_{N} )^{2}$.  When
differentiating the loop function with respect to $w$, the integrand
is positive definite regardless of the value of $s^{\prime} (> s_{\rm
  th})$.  Therefore, the condition that the derivative of the loop
function becomes negative (positive) at the nucleon mass is
(un-)physical.  However, in actual calculations, the positive value
for the derivative of the loop function can happen to appear according
to the value of the subtraction constant $\tilde{A}$.

Based on this discussion, in order to resolve the problem that the
derivative of the loop function becomes positive, in addition to $G_{L
  = 1} ( M_{N} ) = 0$ we further constrain the subtraction constant so
that the derivative of the $L = 1$ loop function should be nonpositive
at the nucleon pole: $d G_{L=1} / d w ( M_{N} ) \le 0$.  With this
additional constraint, we obtain the best fit of the scattering
amplitude to the WI~08 solution as the parameter set ``Constrained'',
whose values are listed in the third column of Table~\ref{tab:1}.
This gives a slightly worse $\chi ^{2}$ value but we cannot see any
clear discrepancy from curves in Fig.~\ref{fig:P33} (dashed line).
Properties of $\Delta (1232)$ and $N (940)$ in the parameter set
Constrained are listed in the fourth and fifth columns of
Table~\ref{tab:2}.  The values of the coupling constants and
compositeness are very similar to the parameter set Naive, except for
the $\pi N$ compositeness for $N (940)$, which now becomes
nonnegative.  The result in the present model indicates that the $N
(940)$ state is, as expected, not described in the $\pi N$ molecular
picture.

Finally, we note that there is ambiguity in calculating the $\pi N$
compositeness $X_{\pi N}$ for $p$-wave resonances from the loop
function with the subtraction scheme and the dimensional
regularization~\eqref{eq:GL_dim}.  Namely, as discussed in
Ref.~\cite{Hyodo:2008xr}, we can consider a shift of the subtraction
constant $\tilde{A}$, which can be compensated by the corresponding
shift of the interaction $V$ so as not to change the full amplitude
$T$.  However, this shift of the subtraction constant can change the
value of $d G_{L = 1} / d w$ and hence that of $X_{\pi N}$, since the
subtraction constant survives when we differentiate $G_{\pi N, \, L =
  1} ( w )$ [see the structure in Eq.~\eqref{eq:GL_dim}].
Nevertheless, if we have a constraint $d G_{L =1} / d w ( M_{N} ) \le
0$, such a shift of the subtraction constant is also constrained and
$d G_{L = 1} / d w$ cannot be close to zero around the $\Delta (1232)$
energy region.  This can be seen from lines in Fig.~\ref{fig:Gloop}.
Namely, if the subtraction constant $\tilde{A}$ could increase
arbitrarily only with the constraint $G_{L = 1} ( M_{N} ) = 0$, the
real part of the loop function in $w \ge M_{N}$ could shift upward in
Fig.~\ref{fig:Gloop} and eventually become flat around $w = 1.2 \gev$,
with which the $X_{\pi N}$ compositeness for $\Delta (1232)$ would be
negligible due to the negligible value of $d G_{L = 1} / d w ( w_{\rm
  pole} )$.  However, in such a case the derivative at $w = M_{N}$, $d
G_{L = 1} / d w ( M_{N} )$, should be largely positive and hence it
should be excluded.  Therefore, in the present formulation, we cannot
arbitrarily shift the value of the $\pi N$ compositeness for $\Delta
(1232)$ without changing the scattering amplitude.  In particular, in
the present calculation $\tilde{A}$ takes its maximal value under the
constraint $d G_{L =1} / d w ( w = M_{N} ) \le 0$, as seen from $d
G_{L =1} / d w ( w = M_{N} ) = 0$ in Fig.~\ref{fig:Gloop}.  As a
consequence, the present calculation would give a minimal value of $|
X_{\pi N} |$ for $\Delta (1232)$ in our approach from the viewpoint of
the shift of the subtraction constant.  In the same manner, $N (940)$
could have a certain positive value of the $\pi N$ compositeness by
the shift of the subtraction constant. We note that such ambiguity
will not take place when we use a usual cut-off scheme for the loop
function rather than the dimensional regularization.  In this
condition, the derivative of the loop function at the nucleon mass
will be definitely negative and nonzero, and hence $X_{\pi N}$ for the
nucleon will be positive and nonzero, say, $0.1$.

In summary, from the precise $\pi N$ scattering amplitude we have
found that, in the real part, the $\pi N$ compositeness is larger than
the elementariness for the $\Delta (1232)$ resonance.  Its imaginary
part is nonnegligible, but the value of $U$ is less than one third.
Therefore, we may conclude that the $\pi N$ component in the $\Delta
(1232)$ resonance is large.  The large real part of the $\pi N$
compositeness and its nonnegligible imaginary part might be the origin
of the large meson cloud effect observed in, {\it e.g.}, the $M1$
transition form factor of the $\gamma ^{\ast} N \to \Delta (1232)$
process in the small momentum transfer region~\cite{Sato:2009de}.  We
mention that we have had two problems on the $\pi N$ compositeness for
the $p$-wave states; one is the negative $\pi N$ compositeness for $N
(940)$ in the naive fitting, and the other is ambiguity due to the
shift of the subtraction constant.  Both originate from the value of
the subtraction constant used in the analysis, and we have discussed
the problems from the viewpoint of the shift of the subtraction
constant and constraint on it at the energy of the nucleon mass.  As a
result, we have shown that in our approach the value of $| X_{\pi N}
|$ for $\Delta (1232)$ cannot be small.

\subsection{The $\bm{N (1535)}$ and $\bm{N (1650)}$ resonances}
\label{sec:3B}

Next we consider the $N (1535)$ and $N (1650)$ resonances, both of
which are $S_{11}$ states in the $\pi N$ scattering, and calculate
their meson--baryon compositeness.  In this study we describe these
resonances in an $s$-wave $\pi N$-$\eta N$-$K \Lambda$-$K \Sigma$
coupled-channels scattering in the chiral unitary approach.  Here we
do not introduce explicit pole terms for $N (1535)$ and $N (1650)$, in
contrast to the case of $\Delta (1232)$, since it is a good starting
point to examine the pciture of dominant meson-baryon components for
them, as they can be discussed with meson-baryon dynamics in $s$ wave.
We regard that missing contributions are implemented into the energy
dependence of the interaction, not as explicit channels coupling to
meson-baryon states.  Nevertheless, the essential part of the
discussion about the elementariness is not changed, as we have done in
the previous section.

\subsubsection{Scattering amplitude}

First we construct the interaction kernel in the chiral unitary
approach.  In this study we take into account the Weinberg--Tomozawa
term [Fig.~\ref{fig:V}(a)] and the next-to-leading order term
[Fig.~\ref{fig:V}(b)] for the interaction kernel, as done in
Ref.~\cite{Bruns:2010sv}.  From the Lagrangian of chiral perturbation
theory, we obtain the interaction before the $s$-wave projection:
\begin{align}
\mathcal{V}_{j k} = & A_{\rm WT} \Slash{R} 
+ A_{\rm M} 
+ A_{14} ( q \cdot q^{\prime} )
+ A_{57} [ \Slash{q} , \, \Slash{q}^{\, \prime} ]
\notag \\
& + A_{811} [ \Slash{q}^{\, \prime} ( q \cdot P ) 
+ \Slash{q} ( q^{\prime} \cdot P ) ] ,
\end{align}
where $j$, $k = \pi N$, $\eta N$, $K \Lambda$, and $K \Sigma$ are the
channel indices, $q^{\mu}$ and $q^{\prime \mu}$ are the meson momenta
in the initial and final states, respectively, $\Slash{q} \equiv
\gamma ^{\mu} q_{\mu}$ with the Dirac gamma matrices $\gamma ^{\mu}$,
$R^{\mu} \equiv q^{\mu} + q^{\prime \mu}$, and $A_{\rm WT}$, $A_{\rm
  M}$, $A_{14}$, $A_{57}$, and $A_{811}$ are the coefficients of the
meson--baryon couplings determined by flavor SU(3) symmetry together
with the low-energy constants, meson decay constants, and meson
masses.  The expression of the coefficients $A_{\rm WT}$, $A_{\rm M}$,
$A_{14}$, $A_{57}$, and $A_{811}$ as well as the pertinent Lagrangian
of chiral perturbation theory can be found in
Ref.~\cite{Bruns:2010sv}.  We have 14 low-energy constants in the
coefficients, $b_{1}$ to $b_{11}$, $b_{0}$, $b_{D}$, and $b_{F}$, and
we treat them as the model parameters.  The meson decay constants are
chosen at their physical values: $f_{\pi} = 92.4 \mev$, $f_{K} = 1.2
f_{\pi}$, and $f_{\eta} = 1.3 f_{\pi}$.  The interaction $\mathcal{V}$
is projected to the $s$ wave as
\begin{equation}
  V_{I=1/2 \, L=0, \, j k}^{+} ( w )
  = \left [ \bar{u}_{j} \mathcal{V}_{j k} u_{k} \right ] _{s\text{-wave}} ,
\end{equation}
where $w$ is the center-of-mass energy, $u_{j}$ is the Dirac spinor
for the $j$th channel baryon, whose normalization is summarized in
Appendix~\ref{app:1}, and $\bar{u}_{j} \equiv u_{j}^{\dagger} \gamma
^{0}$.  The $s$-wave projection of each term can be performed
as\footnote{We note that the term $\bar{u} ( q \cdot q^{\prime} ) u$
  has a higher-order $s$-wave part proportional to $| \bm{q} |^{2} |
  \bm{q}^{\prime} |^{2}$, which is neglected in this study.}
\begin{equation}
  [ \bar{u}_{j} \Slash{R} u_{k} ] _{s\text{-wave}}
  = \mathcal{N}_{j} \mathcal{N}_{k} ( 2 w - M_{j} - M_{k}) ,
\end{equation}
\begin{equation}
  [ \bar{u}_{j} u_{k} ] _{s\text{-wave}}
  = \mathcal{N}_{j} \mathcal{N}_{k} , 
\end{equation}
\begin{equation}
  [ \bar{u}_{j} ( q \cdot q^{\prime} ) u_{k} ] _{s\text{-wave}}
  = \mathcal{N}_{j} \mathcal{N}_{k} \omega _{j} ( w ) \omega _{k} ( w ) , 
\end{equation}
\begin{align}
  & 
  [ \bar{u}_{j} [\Slash{q} , \, \Slash{q}^{\, \prime} ] u_{k} ] _{s\text{-wave}}
  \notag \\
  & = \mathcal{N}_{j} \mathcal{N}_{k} [ 2 \omega _{j} ( w ) \omega _{k} ( w ) 
  - 2 ( w - M_{j} ) ( w - M_{k}) ] ,
\end{align}
\begin{align}
  & 
  [ \bar{u}_{j} [ \Slash{q}^{\, \prime} ( q \cdot P ) 
    + \Slash{q} ( q^{\prime} \cdot P ) ] u_{k} ]_{s\text{-wave}}
  \notag \\
  & = \mathcal{N}_{j} \mathcal{N}_{k} w [ ( w - M_{j} ) \omega _{k} ( w ) 
    + ( w - M_{k} ) \omega _{j} ( w ) ] ,
\end{align}
where $\mathcal{N}_{j}$ is the normalization factor for the Dirac
spinor:
\begin{equation}
  \mathcal{N}_{j} \equiv \sqrt{E_{j} ( w ) + M_{j}} ,
  \quad
  E_{j} ( w ) \equiv \frac{s + M_{j}^{2} - m_{j}^{2}}{2 w} ,
\end{equation}
and $\omega _{j} ( w )$ is the energy of the $j$th channel meson:
\begin{equation}
\omega _{j} ( w ) \equiv \frac{s + m_{j}^{2} - M_{j}^{2}}{2 w} .
\end{equation}
Here, $s \equiv w^{2}$ and $m_{j}$ and $M_{j}$ are the masses of
meson and baryon in $j$th channel, respectively.  We note that the
constructed interaction kernel $V_{I=1/2 \, L=0, \, j k}^{+}$, which
is abbreviated as $V_{j k}$ in the following, is a function only of
the center-of-mass energy $w$; since $L = 0$, we do not need to
factorize the relative momenta $| \bm{q}^{\prime} |^{L} | \bm{q}
|^{L}$ in contrast to the $p$-wave scattering in the previous
subsection.

By using this interaction kernel, the full scattering amplitude $T_{j
  k}$ is a solution of the Lippmann--Schwinger equation in an
algebraic form
\begin{equation}
  T_{j k} ( w ) = V_{j k} ( w )
  + \sum _{l} V_{j l} ( w ) G_{l} ( w ) T_{l k} ( w ) ,
\end{equation}
with the $j$th channel loop function $G_{j}$.  Here we take a
covariant expression for the loop function
\begin{equation}
G_{j} ( w ) = i \int \frac{d^{4} q}{( 2 \pi )^{4}}
\frac{1}{( q^{2} - m_{j}^{2} ) [ (P - q)^{2} - M_{j}^{2} ]} ,
\end{equation}
with $P^{\mu} = ( w, \, \bm{0} )$, and calculate the integral with the
dimensional regularization.  Its expression is shown in
Eq.~\eqref{eqA:Gdim_explicit} in Appendix~\ref{app:3}.  In this study,
in order to fix the subtraction constant in the loop function, we
require the natural renormalization scheme~\cite{Hyodo:2008xr}.
According to the discussion in Ref.~\cite{Hyodo:2008xr}, we introduce
a matching energy scale, at which the full scattering amplitude $T$
coincides with the chiral interaction $V$ for the consistency of the
low-energy theorem with respect to the spontaneous breaking of chiral
symmetry.  We fix the matching energy as the lowest mass of the
``target'' baryons in the scattering~\cite{Hyodo:2008xr}, i.e., the
nucleon mass $M_{N}$:
\begin{equation}
  G_{j} ( w = M_{N} ) = 0 ,
\end{equation}
for every $j$.  With this condition, the loop function $G_{j}$ becomes
physical, i.e., negative in the region $M_{N} \le w \le m_{j} +
M_{j}$, which can exclude explicit pole contributions from the loop
functions~\cite{Hyodo:2008xr}, and fix the loop function without any
model parameter.

\begin{table}[!t]
  \caption{Fitted parameters for the $\pi N$ amplitude $S_{11}$.  The
    $\chi ^{2}$ value divided by the number of degrees of freedom is
    $\chi ^{2} / N_{\rm d.o.f.} = 94.6 / 167$. }
  \label{tab:3}
  \begin{ruledtabular}
    \begin{tabular*}{8.6cm}{@{\extracolsep{\fill}}lc|lc}
      $b_{1}$ [GeV${}^{-1}$] & $0.469$ &
      $b_{8}$ [GeV${}^{-1}$] & $0.523$ 
      \\
      $b_{2}$ [GeV${}^{-1}$] & $-0.048\phantom{-}$ &
      $b_{9}$ [GeV${}^{-1}$] & $-1.246\phantom{-}$ 
      \\
      $b_{3}$ [GeV${}^{-1}$] & $1.244$ &
      $b_{10}$ [GeV${}^{-1}$] & $0.574$ 
      \\
      $b_{4}$ [GeV${}^{-1}$] & $-1.507\phantom{-}$ &
      $b_{11}$ [GeV${}^{-1}$] & $-0.845\phantom{-}$ 
      \\
      $b_{5}$ [GeV${}^{-1}$] & $-1.091\phantom{-}$ &
      $b_{0}$ [GeV${}^{-1}$] & $-5.513\phantom{-}$ 
      \\
      $b_{6}$ [GeV${}^{-1}$] & $-0.722\phantom{-}$ &
      $b_{D}$ [GeV${}^{-1}$] & $1.708$ 
      \\
      $b_{7}$ [GeV${}^{-1}$] & $3.009$ &
      $b_{F}$ [GeV${}^{-1}$] & $2.516$ \\
    \end{tabular*}
  \end{ruledtabular}
\end{table}

\begin{figure}[!t]
  \centering
  \Psfig{8.6cm}{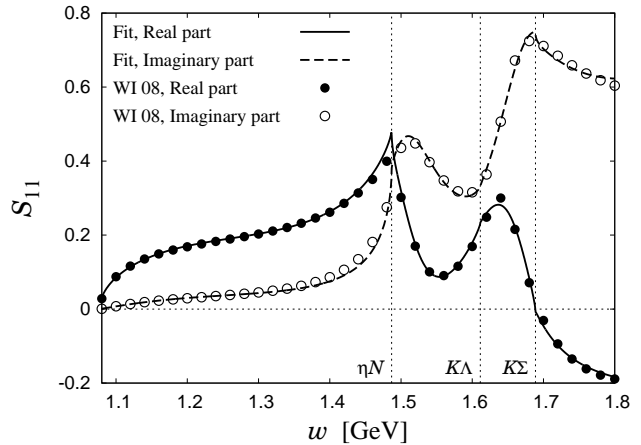}
  \caption{Scattering amplitude $S_{11}$ with parameters fitted to the
    WI~08 solution~\cite{Workman:2012hx}.  The number of the plotted
    data points is $1/5$ of the total in the fits for a better
    visualization.  The vertical dotted lines indicate the $\eta N$,
    $K \Lambda$, and $K \Sigma$ thresholds, respectively.}
  \label{fig:S11}
\end{figure}

Now we can evaluate the scattering amplitude in chiral unitary
approach with 14 model parameters in the interaction kernel.  In the
present calculation, the model parameters are fixed so as to reproduce
the $S_{11}$ solution of the $\pi N$ partial wave analysis
WI~08~\cite{Workman:2012hx}.  As in the previous subsection, we
introduce a normalized scattering amplitude:
\begin{equation}
S_{11} ( w ) 
= - \frac{\rho _{\pi N} (s)}{2} 
T_{\pi N \to \pi N} (w)^{\text{on-shell}} ,
\end{equation}
which satisfies the optical theorem~\eqref{eq:optL} below the
inelastic threshold for the $\pi N$ state.  As we have already
mentioned, the WI~08 solution does not provide errors for the
scattering amplitude.  For the calculation of the $\chi ^{2}$ value,
we introduce errors $0.01$ for $w \le 1.35 \gev$ and $0.02$ for $w >
1.35 \gev$ to the data, which are motivated by the expectation of the
three-body effects above the $\pi \pi N$
threshold~\cite{Nieves:2001wt}.  From the best fit, we obtain the
model parameters listed in Table~\ref{tab:3} with a good value of
$\chi ^{2}$.  We also show the $S_{11}$ amplitude in the theoretical
calculation and the WI~08 solution in Fig.~\ref{fig:S11}, which shows
a good reproduction of the $S_{11}$ amplitude up to $w = 1.8 \gev$.

\subsubsection{Compositeness}

From the $S_{11}$ scattering amplitude, we can search for the
$N^{\ast}$ poles in the complex energy plane.  As a result, we find
two poles located at $w_{\rm pole} = 1496.4 - 58.7 i \mev$ between the
$\eta N$ and $K \Lambda$ thresholds and $w_{\rm pole} = 1660.7 - 70.0
i \mev$ between the $K \Lambda$ and $K \Sigma$ thresholds for the $N
(1535)$ and $N (1650)$ resonances, respectively.  These pole positions
are consistent with the values by Particle Data Group: $w_{\rm pole} =
(1510 \pm 20) - (85 \pm 40) i \mev$ for $N (1535)$ and $w_{\rm pole} =
(1655 \pm 15) - (67.5 \pm 17.5) i \mev$ for $N
(1650)$~\cite{Olive:1900zz}.  In the following we evaluate the
compositeness of the $N (1535)$ and $N (1650)$ resonances by using
this scattering amplitude.  Here we note that, since both resonances
are in the $s$-wave $\pi N$-$\eta N$-$K \Lambda$-$K \Sigma$
coupled-channels scattering, we do not have the problems concerned
with the subtraction constant in the loop function, in contrast to the
case of $\Delta (1232)$ and $N (940)$ in the previous subsection.
Namely, even if we shift the subtraction constant of the $L = 0$ loop
function, it does not change the value of $d G_{L = 0} / d w$ and that
of the compositeness, since the subtraction constant is eliminated
when we perform the derivative of the $L = 0$ loop function.  In other
words, the integral in $d G_{L = 0} / d w$ converges.

According to the scheme developed in Sec.~\ref{sec:2C}, we extract the
compositeness from the scattering amplitude.  From the residue $g_{j}
g_{k}$ of the scattering amplitude at the resonance pole position
$w_{\rm pole}$,
\begin{equation}
T_{j k} ( w ) = \frac{g_{j} g_{k}}{w - w_{\rm pole}} 
+ (\text{regular at } w = w_{\rm pole}) ,
\end{equation}
we evaluate the compositeness as the norm of the two-body wave
function
\begin{equation}
X_{j} = - g_{j}^{2} \frac{d G_{j}}{d w} ( w = w_{\rm pole} ) ,
\end{equation}
and the elementariness as well
\begin{equation}
  Z = - \sum _{j, k} g_{k} g_{j} 
  \left [ 
    G_{j} \frac{d V_{j k}}{d w} G_{k}
  \right ]_{w = w_{\rm pole}} ,
\end{equation}
which measures the contributions from missing channels on the
assumption that the energy dependence of the interaction originates
from missing channels.  The normalization of the total wave function
is achieved as
\begin{equation}
  \sum _{j} X_{j} + Z = 1 .
  \label{eq:sum_rule_Nstar}
\end{equation}
From the compositeness and elementariness, both of which are complex
for resonances in general, we calculate quantities which are real,
bounded in the range $[0, \, 1]$, and automatically satisfy the sum
rule:
\begin{equation}
\tilde{X}_{j} \equiv \frac{ | X_{j} | }{1 + U} , \quad 
\tilde{Z} \equiv \frac{ | Z | }{1 + U} , 
\end{equation}
with 
\begin{equation}
U \equiv \sum _{j} | X_{j} | + | Z | - 1 . 
\end{equation}
Obviously, we have the sum rule for $\tilde{X}_{j}$ and $\tilde{Z}$:
\begin{equation}
\sum _{j} \tilde{X}_{j} + \tilde{Z} = 1 .
\end{equation}
We can interpret $\tilde{X}_{j}$ and $\tilde{Z}$ as the
``probability'' if and only if $U$ is much smaller than unity, $U \ll
1$.

\begin{table}[!t]
  \caption{Properties of $N (1535)$ and $N (1650)$.}
  \label{tab:4}
  \begin{ruledtabular}
    \begin{tabular*}{8.6cm}{@{\extracolsep{\fill}}lcc}
      & $N (1535)$ & $N (1650)$ \\
      \hline
      $w_{\rm pole}$ [MeV] 
      & $1496.4 - 58.7 i$ 
      & $1660.7 - 70.0 i$ 
      \\
      $g_{\pi N}$ [MeV${}^{1/2}$] 
      & $47.1 - \phantom{0}7.3 i$ 
      & $49.8 - 23.1 i$ 
      \\
      $g_{\eta N}$ [MeV${}^{1/2}$] 
      & $68.9 - 42.4 i$ 
      & $-19.0 + 11.1 i\phantom{-}$ 
      \\
      $g_{K \Lambda}$ [MeV${}^{1/2}$] 
      & $85.0 + 14.4 i$ 
      & $-29.9 + 37.1 i\phantom{-}$ 
      \\
      $g_{K \Sigma}$ [MeV${}^{1/2}$] 
      & $-31.4 + 17.5 i\phantom{-}$ 
      & $-73.8 + \phantom{0}6.0 i\phantom{-}$ 
      \\
      $X_{\pi N}$ 
      & $-0.02 + 0.03 i\phantom{-}$ 
      & $0.00 + 0.04 i$ 
      \\
      $X_{\eta N}$ 
      & $0.04 + 0.37 i$ 
      & $0.00 + 0.01 i$ 
      \\
      $X_{K \Lambda}$ 
      & $0.14 + 0.00 i$ 
      & $0.08 + 0.05 i$ 
      \\
      $X_{K \Sigma}$ 
      & $0.01 - 0.02 i$ 
      & $0.09 - 0.12 i$ 
      \\
      $Z$ 
      & $0.84 - 0.38 i$ 
      & $0.84 + 0.01 i$ 
      \\
      $U$ & $0.48$ & $0.13$ \\
      $\tilde{X}_{\pi N}$ & $0.03$ & $0.04$ \\
      $\tilde{X}_{\eta N}$ & $0.25$ & $0.01$ \\
      $\tilde{X}_{K \Lambda}$ & $0.09$ & $0.08$ \\
      $\tilde{X}_{K \Sigma}$ & $0.01$ & $0.13$ \\
      $\tilde{Z}$ & $0.62$ & $0.74$ 
    \end{tabular*}
  \end{ruledtabular}
\end{table}

The numerical results of the coupling constants, compositeness, and
elementariness are listed in Table~\ref{tab:4} both for the $N (1535)$
and $N (1650)$ resonances.

For the $N (1535)$ resonance, its coupling constants show an ordering
similar to that in Ref.~\cite{Bruns:2010sv}; in particular, $| g_{K
  \Lambda} |$ is the largest and $| g_{\eta N} |$ comes next, which is
consistent with the result in Ref.~\cite{Bruns:2010sv}.  However, the
values of the compositeness in the $K \Lambda$ and $\eta N$ channels
are not comparable to unity, and the elementariness $Z$ dominates the
sum rule~\eqref{eq:sum_rule_Nstar}.  Therefore, our result implies
that $N (1535)$ has a large component originating from contributions
other than the pseudoscalar meson--baryon dynamics considered.  This
conclusion was already drawn in Ref.~\cite{Sekihara:2014kya} with the
simplest interaction, i.e., the Weinberg--Tomozawa term, and we
confirm this with our refined model for the precise $S_{11}$
amplitude.  The result of the compositeness means that the
missing-channel contribution $Z$ dominates the sum rule even if we do
not take into account a bare-state contribution explicitly.  The
missing channel can contribute to the appearance of the resonance
through the energy dependence of the interaction and the low-energy
constants.  In other words, in the present framework, information on
the $N (1535)$ resonance is encoded in the energy dependence of the
chiral interaction and the low-energy constants in it.  However, in
the present model space, we cannot conclude what the missing channel
is; we expect that this will be genuine one-body state, but other
channels such as vector meson-baryon and meson-meson-baryon systems
could be origin.\footnote{If we can reproduce well the $S_{11}$
  amplitude with the two bare pole terms corresponding to $N (1535)$
  and $N (1650)$ and energy independent meson--baryon interaction, we
  can conclude that $N (1535)$ and $N (1650)$ originate from one-body
  states, respectively.}  We also note that the value of $U$ is not
small compared to unity, due to the nonnegligible imaginary part of
$\tilde{X}_{\eta N}$ and $\tilde{Z}$.  Therefore, modified quantities
$\tilde{X}_{j}$ and $\tilde{Z}$ cannot be interpreted as probabilities
to find the composite and missing fractions, respectively.  In
particular, although $\tilde{X}_{\eta N}$ is one fourth, we cannot
conclude a nonnegligible $\eta N$ component for $N (1535)$.

Next, for the $N (1650)$ resonance, $| g_{K \Sigma} |$ is the largest
among the absolute values of the coupling constants, as in
Refs.~\cite{Bruns:2010sv}.  However, the ordering of the coupling
constants is not consistent.  We expect that this is mainly because
the accuracy of the fitting.  Actually, our fitting can be more
accurate, as seen in the better reproduction of the $N (1650)$ pole
position reported by Particle Data Group.  As for the component of $N
(1650)$, we can see that the elementariness $Z$ dominates the sum
rule~\eqref{eq:sum_rule_Nstar}.  In addition, the value of $U$ for
$N(1650)$ is much smaller than unity.  Therefore, we can safely
interpret the modified quantities $\tilde{X}_{j}$ and $\tilde{Z}$ as
probabilities.  The result listed in Table~\ref{tab:4} indicates that
$\tilde{Z}$ is dominant and hence the $N (1650)$ resonance is indeed
dominated by contributions other than the pseudoscalar meson--baryon
dynamics considered.

Finally it is interesting to compare the structure of $N (1535)$ and
$N (1650)$ with that of $\Lambda (1405)$ and $\Xi (1690)$, all of
which are considered to be dynamically generated in the chiral unitary
approach.  The compositeness of $\Lambda (1405)$ was evaluated in the
chiral unitary approach in Ref.~\cite{Sekihara:2014kya} with the
leading plus next-to-leading order chiral
interaction~\cite{Ikeda:2011pi, Ikeda:2012au}, concluding that the
higher pole of $\Lambda (1405)$ is indeed dominated by the $\bar{K} N$
composite state.  In contrast to $\Lambda (1405)$, the compositeness
of $N (1535)$ and $N (1650)$ is negligible or not large, although we
describe $N (1535)$ and $N (1650)$ in the meson--baryon degrees of
freedom, as in the $\Lambda (1405)$ case.  This difference of the
structure is expected to originate from the different thresholds and
model parameters (low-energy constants and subtraction constants),
both of which should degenerate in the SU(3) symmetric world.  In
particular, when we shift the system from the SU(3) symmetric world to
the physical one, the situation in the $S = 0$ sector would change
most drastically; the $\pi N$ threshold becomes the lowest one and the
other channels such as $\rho N$, $\pi \Delta$, and genuine $q q q$
states would contribute to the $\pi N$ scattering.  In this study
these are reflected to the low-energy constants in the next-to-leading
order as the missing channels.  Actually, while the chiral unitary
approach can reproduce the phenomena around the $\bar{K} N$ threshold
for $\Lambda (1405)$ even with the simplest interaction, i.e., the
Weinberg--Tomozawa interaction~\cite{Ikeda:2011pi, Ikeda:2012au}, the
$\pi N$ scattering amplitude cannot be reproduced well in the chiral
unitary approach around the $N^{\ast}$ region only with the
Weinberg--Tomozawa interaction~\cite{Inoue:2001ip}.  Significant
contributions in the next-to-leading order can introduce missing
channels through the low-energy constants, and hence the compositeness
(elementariness) is small (large) for $N (1535)$ and $N (1650)$.
Besides, we also mention the fate of the dynamically generated
resonances in $S = 0$ and $S = -2$ channels in the chiral unitary
approach.  Interestingly, the Clebsch--Gordan coefficients for the
Weinberg--Tomozawa interaction term are the same for $S = 0$ ($\pi N$,
$\eta N$, $K \Lambda$, $K \Sigma$) and $S = -2$ ($\pi \Xi$, $\eta
\Xi$, $\bar{K} \Lambda$, $\bar{K} \Sigma$) channels.  On the one hand,
in $S = -2$ channel, it was suggested in Ref.~\cite{Sekihara:2015qqa}
that $\Xi (1690)$ can be dynamically generated near the $\bar{K}
\Sigma$ threshold with a dominant $\bar{K} \Sigma$ compositeness in
$s$ wave in the chiral unitary approach, consistently with the
experimental observations.  On the other hand, such a dynamically
generated $N^{\ast}$ state would exist if the Weinberg--Tomozawa
interaction were dominant in the $S = 0$ chiral unitary approach, but
in fact no hadronic molecular $N^{\ast}$ state appears in $S = 0$ due
to the significant contributions from the next-to-leading order terms.
Nevertheless, we think that, although the internal structure of $N
(1535)$, $\Lambda (1405)$, and $\Xi (1690)$ is different, our result
does not mean that they are not members of the same flavor SU(3)
multiplet.  Namely, when we consider the SU(3) multiplets, we measure
the masses of hadrons from the vacuum, that is, from zero.  On the
other hand, when we consider the internal structure in terms of the
compositeness, we measure the masses from the thresholds.  As a
consequence, the breaking effect of flavor SU(3) symmetry is
nonnegligible in terms of the compositeness, since the meson-baryon
thresholds are different for each state.

In summary, we have evaluated the compositeness of the $N (1535)$ and
$N (1650)$ resonances from the precise $S_{11}$ scattering amplitude.
The results indicate that both of them are dominated by components
other than the pseudoscalar meson--baryon dynamics considered.  An
important finding is that the missing-channel contribution $Z$
dominates the sum rule even if we do not take into account a
bare-state contribution explicitly.  The missing channel can
contribute to the appearance of the resonance through the energy
dependence of the interaction.  Finally we note that we do not have
problems concerned with the subtraction constant in the loop function,
in contrast to the case of $\Delta (1232)$ and $N (940)$ in the
previous subsection, since $N (1535)$ and $N (1650)$ are in the
$s$-wave $\pi N$-$\eta N$-$K \Lambda$-$K \Sigma$ coupled-channels
scattering and hence $d G_{L = 0} / d w$ converges.

\section{Summary and outlook}
\label{sec:4}

In this study we have presented a formulation of the compositeness for
baryonic resonances in order to discuss the meson--baryon molecular
structure inside the resonances.  For this purpose, we have shown that
the residue of the scattering amplitude at the resonance pole position
contains the wave function of the resonance with respect to the
two-body channel, both in the nonrelativistic and relativistic
formulations.  Then, we have defined the compositeness for the
resonance state as a norm of the two-body wave function extracted from
the residue of the scattering amplitude.  An important point to be
noted is that the value of compositeness, i.e., the norm of the
two-body wave function, is automatically fixed when we calculate the
residue of the scattering amplitude, without normalizing the wave
function by hand.  We have also defined the missing-channel
contribution, which we call elementariness, as unity minus the sum of
the compositeness, which measures the contributions from missing
channels on the assumption that the energy dependence of the
interaction originates from missing channels.  In addition, from the
compositeness and elementariness, we have introduced quantities which
are real, bounded in the range $[0, \, 1]$, and automatically satisfy
the sum rule.  These quantities can be interpreted as probabilities in
a certain class of resonances.

The formulated compositeness and elementariness were applied to the
$\Delta (1232)$, $N (1535)$, and $N (1650)$ resonances and $N (940)$
in the chiral unitary approach, since there are several implications
that these resonances may have certain fractions of the meson--baryon
components.  In the present model, we have determined the separable
interaction of the pseudoscalar meson--octet baryon from chiral
perturbation theory up to the next-to-leading order.  The $\Delta
(1232)$ resonance and $N (940)$ were described in the $\pi N$
single-channel scattering, while the $N (1535)$ and $N (1650)$
resonances were in the $s$-wave $\pi N$-$\eta N$-$K \Lambda$-$K
\Sigma$ coupled-channels scattering.  In both cases, the loop function
was evaluated with the subtraction scheme and the dimensional
regularization.  In particular, we have to introduce two subtraction
constants for $\Delta (1232)$ and $N (940)$ in order to calculate
their compositeness in a proper way.  The model parameters were fixed
so that the $\pi N$ scattering amplitude precisely reproduces the
solution of the partial wave analysis.

As a result for $\Delta (1232)$, we have found that the real part of
the $\pi N$ compositeness is larger than the elementariness.  The
imaginary part of the $\pi N$ compositeness for $\Delta (1232)$ is
nonnegligible, but the sum of the absolute values of the compositeness
and elementariness is close to unity.  Therefore, we may conclude that
the $\pi N$ component in $\Delta (1232)$ is significant.  We have also
had two problems on the $\pi N$ compositeness for the $p$-wave states;
one is the negative $\pi N$ compositeness for $N (940)$ in the naive
fitting, and the other is ambiguity due to the shift of the
subtraction constant.  Both originate from the value of the
subtraction constant used in the analysis, and we have discussed the
problems from the viewpoint of the shift of the subtraction constant
and constraint on it at the energy of the nucleon mass.  As a
consequence, we have shown that in our approach the absolute value of
the $\pi N$ compositeness for $\Delta (1232)$ cannot be small.

For $N (1535)$ and $N (1650)$, on the other hand, we have found that
both of them are dominated by components other than the pseudoscalar
meson--baryon dynamics considered.  An important finding was that,
even if we do not take into account a bare pole term for the resonance
explicitly, a missing channel can contribute to the appearance of the
resonance through the energy dependence of the interaction and the
low-energy constants.  Since both resonances are in the $s$-wave $\pi
N$-$\eta N$-$K \Lambda$-$K \Sigma$ coupled-channels scattering, we do
not have problems concerned with the subtraction constant in the loop
function, in contrast to the case of $\Delta (1232)$ and $N (940)$.

Finally, we mention that the large absolute value of the $\pi N$
compositeness for the $\Delta (1232)$ resonance would lead to the
large meson cloud effect observed in, {\it e.g.}, the $M1$ transition
form factor of the $\gamma ^{\ast} N \to \Delta (1232)$ process in the
small momentum transfer region.  However, our result relies on the
separable interaction in the form of Eq.~\eqref{eq:separable_Delta},
which might be a too much simplified interaction in describing the
$\pi N$ scattering amplitude especially in the $\Delta (1232)$
resonance region.  In this sense, it would be better to evaluate the
compositeness for $\Delta (1232)$ in solving the integral equation for
the scattering amplitude, such as in the dynamical approaches, so as to
conclude the $\pi N$ structure of $\Delta (1232)$ more clearly.

\begin{acknowledgments}
  The authors acknowledge A.~Hosaka, H.~Nagahiro, T.~Hyodo, and
  D.~Jido for fruitful discussions on the interpretation of the
  compositeness, A.~Dot\'e on the Gamow vectors for resonance
  states, and H.~Kamano on the $N^{\ast}$ resonances in the
  $\pi N$ scattering amplitude.
  This work is partly supported by the Grants-in-Aid for Scientific
  Research from MEXT and JSPS (No.~15K17649, 
  No.~15J06538
  ).
\end{acknowledgments}

\appendix

\section{Conventions}
\label{app:1}

In this Appendix we summarize our conventions of meson--baryon
scatterings used in this paper.  

Throughout this paper we employ the metric in four-dimensional
Minkowski space defined as $g^{\mu \nu} = g_{\mu \nu} = \text{diag}(1,
\, -1, \, -1, \, -1)$ and the Einstein summation convention is used
for the Lorentz index $\mu$.  The Dirac matrices $\gamma ^{\mu}$ ($\mu
= 0$, $1$, $2$, $3$) satisfies the anticommutation relation:
\begin{equation}
\{ \gamma ^{\mu} , \gamma ^{\nu} \} = 2 g^{\mu \nu} . 
\end{equation}
The Dirac spinor $u ( \bm{p}, \, s)$, where $\bm{p}$ is three-momentum
of the field and $s$ represents its spin, is introduced as the
positive energy solution of the Dirac equation for baryons.  In this
paper the Dirac spinor is normalized as follows:
\begin{equation}
\overline{u}(\bm{p}, \, s^{\prime}) u(\bm{p}, \, s) 
= 2 M \delta _{s s^{\prime}} , 
\end{equation}
where $\overline{u} \equiv u^{\dagger} \gamma ^{0}$ and $M$ is the
mass of the Dirac field.

In order to describe the meson--baryon scatterings, we introduce meson
and baryon one-particle states.  The meson states $| \bm{k}, \, j
\rangle$ are normalized in the following manner:
\begin{equation}
\langle \bm{k}^{\prime}, \, j^{\prime} | 
\bm{k}, \, j \rangle 
= 2 \omega _{j} (k) ( 2 \pi )^{3} \delta ^{3} ( \bm{k}^{\prime} - \bm{k} ) 
\delta _{j j^{\prime}} ,
\end{equation}
where $\bm{k}$ is the three-momentum of the meson, $j$ indicates the
channel, and $\omega _{j} (k) \equiv \sqrt{k^{2} + m_{j}^{2}}$ is the
meson energy with $k \equiv | \bm{k} |$.  The baryon states $| \bm{p},
\, s , \, j \rangle$, on the other hand, are normalized in the
following manner:
\begin{equation}
\langle \bm{p}^{\prime}, \, s^{\prime} , \, j^{\prime} | 
\bm{p}, \, s , \, j \rangle 
= 2 E_{j} (p) ( 2 \pi )^{3} \delta ^{3} ( \bm{p}^{\prime} - \bm{p} ) 
\delta _{s s^{\prime}} \delta _{j j^{\prime}} ,
\end{equation}
where $\bm{p}$ is the three-momentum of the baryon, $s$ is its spin,
and $E_{j} (p) \equiv \sqrt{p^{2} + M_{j}^{2}}$.  By using these
states, we can compose $j$th channel meson--baryon two-body state in
the center-of-mass frame with the relative momentum $\bm{q}$, which we
simply write as $| j \rangle$:
\begin{equation}
| j \rangle \equiv 
| \bm{q}, \, j \rangle \otimes | - \bm{q}, \, s , \, j \rangle .
\end{equation}
Then we can evaluate the scattering amplitude $\mathcal{T}_{j k}$ as
the matrix element of the $T$-matrix with respect to the scattering
states $| j \rangle$:
\begin{equation}
\langle j | \hat{T} | k \rangle = 
\bar{u}_{j} \mathcal{T}_{j k} u_{k} , 
\end{equation}
where $u_{j}$ is the Dirac spinor for the $j$th channel baryon.  Note
that we do not sum the channel components $j$ and $k$.  In this
Appendix, characters in the script style denotes four times four
matrices which are sandwiched by the Dirac spinors, except for
Lagrangians.

The scattering amplitude $\bar{u}_{j} \mathcal{T}_{j k} u_{k}$ can be
decomposed into the partial wave amplitudes.  For this purpose we
write the scattering amplitude in the center-of-mass frame as
\begin{align}
& \langle j | \mathcal{T} | k \rangle 
= \bar{u}_{j} \mathcal{T}_{j k} u_{k} 
\notag \\
& = \chi _{j}^{\dag} \left [ 
g_{j k} ( w , \, x) 
- i h_{j k} ( w , \, x) ( \hat{q}_{j} \times \hat{q}_{k} )
\cdot \bm{\sigma} \right ] \chi _{k} , 
\end{align}
where $\chi _{j}$ is the Pauli spinor for the $j$th channel baryon, $w
\equiv \omega _{j} + E_{j} = \omega _{k} + E_{k}$ is the
center-of-mass energy, $x = \cos \theta$ with the center-of-mass
scattering angle $\theta$, $\hat{q}_{j}$ is the unit vector in the
direction of the relative three-momentum in the $j$th channel.  In our
notation we can calculate the differential cross section of the
meson--baryon scatterings in the center-of-mass frame as
\begin{equation}
\frac{d \sigma _{k \to j}}{d \Omega} 
= \frac{1}{64 \pi ^{2} s} \frac{q_{j}}{q_{k}}
\left [ | g_{j k} ( w , \, x ) |^{2} 
+ | h_{j k} ( w , \, x ) |^{2}  \sin ^{2} \theta \right ] ,
\end{equation}
where $s \equiv w^{2}$ and $q_{j}$ is the $j$-channel center-of-mass
momenta.  Then $g_{j k}$ and $h_{j k}$ are expressed in terms of the
partial wave amplitudes $T_{L}^{\pm} (w)$ as
\begin{equation}
g_{j k} ( w , \, x) = \sum _{L=0}^{\infty} 
\left [ ( L + 1 ) T_{L}^{+} ( w ) + L T_{L}^{-} ( w ) 
\right ]_{j k} P_{L} ( x )  ,
\end{equation}
\begin{equation}
h_{j k} ( w , \, x) = \sum _{L=1}^{\infty} 
\left [ T_{L}^{+} ( w ) - T_{L}^{-} ( w ) 
\right ]_{j k} P_{L}^{\prime} ( x ) , 
\end{equation}
with the Legendre polynomials $P_{L} (x)$ and $P_{L}^{\prime} (x)
\equiv d P_{L} / d x$.  In terms of $T_{L}^{\pm} (w)$ the optical
theorem can be expressed as
\begin{equation}
\text{Im} \, \left [ T_{L}^{\pm} ( w ) \right ]_{j j} 
= - \frac{1}{8 \pi w} \sum _{k} q_{k} 
\left | \left [ T_{L}^{\pm} ( w ) \right ]_{j k} \right | ^{2} ,
\end{equation}
where the sum runs over the open channels.

The partial wave amplitude $T_{L}^{\pm} (w)$ can be extracted from the
scattering amplitudes $g_{j k}$ and $h_{j k}$ by using the orthonormal
relation
\begin{equation}
\int _{-1}^{1} d x P_{L} ( x ) P_{M} ( x ) 
= \frac{2}{2 L + 1} \delta _{L M} , 
\end{equation}
and a relation
\begin{equation}
\int _{-1}^{1} d x P_{L}^{\prime} ( x ) [ P_{M \pm 1} ( x ) - x P_{M} ( x ) ]
= \left ( \frac{1}{2 L + 1} \mp 1 \right ) \delta _{L M} . 
\end{equation}
Actually, from these relations we can extract $T_{L}^{\pm}$ as
\begin{align}
& T_{L}^{\pm} ( w )_{j k} 
\notag \\
& = \frac{1}{2} \int _{-1}^{1} d x 
\left [ g_{j k} P_{L} ( x ) 
- h_{j k} ( P_{L \pm 1} (x) - x P_{L} (x) ) \right ] .
\end{align}

It is useful to express the meson--baryon scattering amplitude
$\mathcal{T}$ in the following way:
\begin{align}
\mathcal{T}
= A ( s, \, t) 
+ \frac{\Slash{R}}{2} 
B ( s , \, t) , 
\label{eqA:AB}
\end{align}
where $s$ and $t$ are Mandelstam variables, $\Slash{p} \equiv \gamma
^{\mu} p_{\mu}$, and $R^{\mu} \equiv q^{\mu} + q^{\prime \mu}$ with
$q^{( \prime ) \mu}$ being the meson momentum in the initial (final)
state.  In particular, in the single-channel problem, from $A (s, \,
t)$ and $B (s, \, t)$ we can calculate the scattering amplitudes $g (
w , \, x )$ and $h ( w , \, x )$ as
\begin{align}
g = & \frac{1}{2 ( E + M )}
\big \{ 
[ 4 M ( E + M ) - t ] A 
\notag \\
& + [ ( w + M ) t + 4 ( E + M ) ( w E - M^2 ) ] B 
\big \} ,
\end{align}
\begin{equation}
h = \frac{A - ( w + M) B}{E + M} | \bm{q} |^{2} ,
\end{equation}
where $M$ and $E$ are mass and energy of the baryon, respectively.

Finally we mention that the $\pi N$ scattering amplitude used in
Sec.~\ref{sec:3}, which we denote as $L_{2 I \, 2 J}$ with isospin $I$
and total angular momentum $J = L \pm 1/2$, is expressed with the
scattering amplitude $T_{L}^{\pm}$ in our convention as
\begin{equation}
L_{2 I \, 2 J} ( w ) = - \frac{q_{\pi N}}{8 \pi w} T_{L}^{\pm} (w).
\end{equation}
Therefore, below the inelastic threshold the $\pi N$ scattering
amplitude $L_{2 I \, 2 J}$ satisfies the optical theorem
\begin{equation}
\text{Im} \, L_{2 I \, 2 J} = | L_{2 I \, 2 J} |^{2} ,
\end{equation}
for each partial wave.

\section{$\bm{\pi N}$ interaction kernel from chiral
  perturbation theory}
\label{app:2}

In this Appendix we show the expressions of the $\pi N$ interaction
kernel used in Sec.~\ref{sec:3A}.  In general, the $\pi N$ scattering
has two isospin components, namely, $I=3/2$ and $I=1/2$, respectively.
The $\pi N (I=3/2)$ amplitude corresponds to the $\pi ^{+} p \to \pi
^{+} p$ one:
\begin{equation}
\mathcal{T}_{\pi N (I=3/2)} = \mathcal{T}_{\pi ^{+} p \to \pi ^{+} p} .
\end{equation}
On the other hand, the $\pi N (I=1/2)$ amplitude can be written in
terms of the $\pi ^{-} p \to \pi ^{-} p$ and $\pi ^{+} p \to \pi ^{+}
p$ amplitudes:
\begin{equation}
\mathcal{T}_{\pi N (I=1/2)} 
= \frac{3}{2} \mathcal{T}_{\pi ^{-} p \to \pi ^{-} p} 
- \frac{1}{2} \mathcal{T}_{\pi ^{+} p \to \pi ^{+} p} .
\end{equation}
By using crossing symmetry, we can convert the $\pi ^{+} p \to \pi
^{+} p$ amplitude into $\pi ^{-} p \to \pi ^{-} p$ one in terms of $A$
and $B$ in Eq.~\eqref{eqA:AB}:
\begin{equation}
A_{\pi ^{-} p \to \pi ^{-} p} ( s , \, t ) =  
A_{\pi ^{+} p \to \pi ^{+} p} ( u , \, t ) , 
\end{equation}
\begin{equation}
B_{\pi ^{-} p \to \pi ^{-} p} ( s , \, t ) =  
- B_{\pi ^{+} p \to \pi ^{+} p} ( u , \, t ) . 
\end{equation}
Therefore, in the following we consider only the $\pi ^{+} p \to \pi
^{+} p$ amplitude.

In this study we employ chiral perturbation theory up to $\mathcal{O}
(p^{2})$ for the $\pi ^{+} p \to \pi ^{+} p$ interaction kernel
$\mathcal{V}$.  The interaction kernel consists of the
Weinberg--Tomozawa term $\mathcal{V}_{\rm WT}$, $u$-channel $N(940)$
exchange $\mathcal{V}_{s+u}$, next-to-leading order contact term
$\mathcal{V}_{2}$, and $s$- and $u$-channel $\Delta (1232)$ exchanges
$\mathcal{V}_{\Delta}$:
\begin{equation}
  \mathcal{V} = \mathcal{V}_{\rm WT} + \mathcal{V}_{s+u} 
  + \mathcal{V}_{2} + \mathcal{V}_{\Delta} .
\end{equation}
The interaction kernel $\mathcal{V}$ in the isospin basis is projected
to the partial wave amplitude $V_{I L}^{\pm} (w)$ in the same way as
$\mathcal{T} \to T_{I L}^{\pm} (w)$ shown in Appendix~\ref{app:1}.
This partial wave amplitude corresponds to the interaction kernel in
Eq.~\eqref{eq:Vproj} and then is unitarized as in Sec.~\ref{sec:3A}.

The leading order [$\mathcal{O} (p^{1})$] pion--nucleon Lagrangian can
be expressed as
\begin{equation}
\mathcal{L}_{\pi N}^{(1)} = \bar{N} ( i \Slash{D} - M ) N
+ \frac{g}{2} \bar{N} \Slash{u} \gamma _{5} N .
\end{equation}
In the expression, $N = (p, \, n)^{t}$ is the nucleon fields, $M$ and
$g$ are the nucleon mass and the nucleon axial charge in the chiral
limit, respectively, and $D_{\mu} \equiv \prt _{\mu} + \Gamma _{\mu}$
is the covariant derivative with $\Gamma _{\mu} = [u^{\dagger}, \,
\prt _{\mu} u]/2$, where $u$ is the square root of $U$ in the
nonlinear representation:
\begin{align}
u (x) \equiv & \sqrt{U (x)} 
= \exp \left [ i \frac{\vec{\pi} (x) \cdot \vec{\tau}}{2 f} \right ] 
\notag \\
= & 1 + i \frac{\vec{\pi} (x) \cdot \vec{\tau}}{2 f} 
- \frac{\vec{\pi} (x)^{2}}{8 f^{2}} 
+ \mathcal{O} ( ( \vec{\pi} \cdot \vec{\tau} )^{3}) ,
\end{align}
with the pion decay constant $f$ in the chiral limit and $\vec{\tau}$
being the Pauli matrices acting in isospin space.  The pion fields
$\vec{\pi}$ are expressed as:
\begin{equation}
\vec{\pi} (x) \cdot \vec{\tau} = 
  \begin{pmatrix}
    \pi ^{0} (x) & \sqrt{2} \pi ^{+} (x) \\
    \sqrt{2} \pi ^{-} (x) & - \pi ^{0} (x) \\
  \end{pmatrix} ,
\end{equation}
and we further define $u_{\mu}$ as
\begin{equation}
u_{\mu} \equiv i u^{\dagger} \prt _{\mu} U u^{\dagger} 
= - \frac{\prt _{\mu} \vec{\pi} (x) \cdot \vec{\tau}}{f} 
+ \mathcal{O} ( ( \vec{\pi} \cdot \vec{\tau} )^{3}) .
\end{equation}
By using them we obtain the leading order $\pi N$ interaction as
\begin{equation}
\mathcal{V}_{\rm WT} = \frac{\Slash{R}}{2} \frac{1}{2 f_{\pi}^{2}} ,
\end{equation}
\begin{equation}
\mathcal{V}_{s + u} = - \frac{g_{A}^{2}}{f_{\pi}^{2}} M_{N} 
+ \frac{\Slash{R}}{2} \left [ - \frac{g_{A}^{2}}{2 f_{\pi}^{2}} 
\frac{u + 3 M_{N}^{2}}{u - M_{N}^{2}} \right ],
\end{equation}
where we have replaced $f$, $M$, and $g$ with their physical values
$f_{\pi}$, $M_{N}$, and $g_{A}$, respectively.  In this study we fix
$f_{\pi} = 92.4 \mev$, $M_{N} = 938.92 \mev$, and $g_{A} = 1.267$.
Therefore, the leading order term does not have model parameters.

The next-to-leading order [$\mathcal{O} (p^{2})$] pion--nucleon
Lagrangian can be expressed, after neglecting irrelevant terms for the
$\pi N$ scattering, as
\begin{align}
\mathcal{L}_{\pi N}^{(2)} = & c_{1} \langle \chi _{+} \rangle \bar{N} N
- \frac{c_{2}}{4 M^{2}} \langle u_{\mu} u_{\nu} \rangle 
\left [ \bar{N} D^{\mu} D^{\nu} N + ( \text{h.c.} ) \right ]
\notag \\
& + \frac{c_{3}}{2} \langle u_{\mu} u^{\mu} \rangle \bar{N} N
- \frac{c_{4}}{4} \bar{N} \gamma ^{\mu} \gamma ^{\nu} [ u_{\mu}, \, u_{\nu} ] N
+ \cdots ,
\end{align}
where $\langle A \rangle$ denotes the trace of the $2 \times 2$
matrix $A$ in the flavor space, and
\begin{equation}
\chi _{+} \equiv u^{\dagger} \chi u^{\dagger} + u \chi ^{\dagger} u ,
\end{equation}
with $\chi = 2 B_{0} \hat{m}$ ($\hat{m} = m_{u} = m_{d}$ in the
isospin symmetric limit), which can be replaced with the squared pion
mass: $\chi \approx m_{\pi}^{2}$.  By using them we obtain the
next-to-leading order $\pi N$ interaction as
\begin{align}
\mathcal{V}_{2} = & \frac{4 c_{1} m_{\pi}^{2}}{f_{\pi}^{2}} 
- \frac{c_{2}}{8 f_{\pi}^{2} M_{N}^{2}} [ ( s - u )^{2} - t^{2} ]
- \frac{c_{3} ( 2 m_{\pi}^{2} - t )}{f_{\pi}^{2}}
\notag \\
& - \frac{c_{4}}{2 f_{\pi}^{2}} ( s - u ) + \frac{\Slash{R}}{2} 
\frac{2 c_{4} M_{N}}{f_{\pi}^{2}} 
.
\end{align}
The mass of pion is fixed as $m_{\pi} = 138.04 \mev$.  The low-energy
constants $c_{1}$, $c_{2}$, $c_{3}$, and $c_{4}$ are model parameters
to be fixed so that we reproduce the scattering amplitude obtained in
the partial wave analysis.

The $s$- and $u$-channel $\Delta (1232)$ exchange term,
$\mathcal{V}_{\Delta}$, is calculated with the
Lagrangian~\cite{Scherer:2012xha}:
\begin{equation}
\mathcal{L}_{\pi N \Delta} = 
\frac{g_{\pi N \Delta}}{m_{\pi}} \bar{\Delta}^{\mu} 
\vec{T}^{\dagger} ( g_{\mu \nu} - \gamma _{\mu} \gamma _{\nu} )
N \prt ^{\nu} \vec{\pi} + ( \text{h.c.} ) ,
\end{equation}
with the bare $\pi N \Delta$ coupling constant $g_{\pi N \Delta}$ and
the $1/2 \to 3/2$ isospin transition operator $\vec{T}$, which
satisfies
\begin{equation}
T_{b} 
T_{a}^{\dagger} 
= \delta _{b a} - \frac{1}{3} \tau _{b} \tau _{a} .
\end{equation}
The spin-$3/2$ propagator with the momentum $P^{\mu}$ is:
\begin{equation}
- i \frac{\Slash{P} + M_{\Delta}}{P^{2} - M_{\Delta}^{2}}
\left ( g_{\mu \nu} - \frac{1}{3} \gamma _{\mu} \gamma _{\nu} 
- \frac{2 P_{\mu} P_{\nu}}{3 M_{\Delta}^{2}} 
+ \frac{P_{\mu} \gamma _{\nu} - P_{\nu} \gamma _{\mu}}{3 M_{\Delta}} 
\right ) ,
\end{equation}
with the bare $\Delta$ mass $M_{\Delta}$.  Then the interaction
$\mathcal{V}_{\Delta}$ is expressed as
\begin{align}
\mathcal{V}_{\Delta} = & A_{\Delta} ( s , \, t )
+ \frac{1}{3} A_{\Delta} ( u , \, t )
\notag \\ & 
+ \frac{\Slash{R}}{2} 
\left [ B_{\Delta} ( s , \, t )
- \frac{1}{3} B_{\Delta} ( u , \, t ) \right ] ,
\end{align}
with
\begin{align}
& A_{\Delta} ( s , \, t ) = 
- \frac{g_{\pi N \Delta}^{2}}{6 M_{\Delta}^{2} m_{\pi}^{2} ( s - M_{\Delta}^{2} )}
\notag \\
& \times 
\big \{ 
3 M_{\Delta}^{2} ( M_{\Delta} + M_{N} )
( 2 M_{N}^{2} - 2 s - t + 2 m_{\pi}^{2} ) 
\notag \\
& \phantom{\times \big \{}
+ 2 M_{\Delta} [ M_{N}^{2} ( m_{\pi}^{2} - 2 s) + 2 s^{2} 
- m_{\pi}^{2} s - m_{\pi}^{4} ]
\notag \\
& \phantom{\times \big \{}
- M_{N} [ M_{N}^{4} - 2 M_{N}^{2} ( m_{\pi}^{2} - s ) 
- 3 s^{2} + 2 m_{\pi}^{2} s + m_{\pi}^{4} ] \big \} ,
\end{align}
\begin{align}
& B_{\Delta} ( s , \, t ) = 
- \frac{g_{\pi N \Delta}^{2}}{6 M_{\Delta}^{2} m_{\pi}^{2} ( s - M_{\Delta}^{2} )}
\big \{ 
12 M_{\Delta}^{3} M_{N} 
\notag \\
& \phantom{\times \big \{}
+ 3 M_{\Delta}^{2} ( 4 M_{N}^{2} - t )
+ 2 M_{\Delta} M_{N} ( M_{N}^{2} - m_{\pi}^{2} - 5 s) 
- M_{N}^{4} 
\notag \\
& \phantom{\times \big \{}
+ 2 M_{N}^{2} ( m_{\pi}^{2} - 3 s ) - ( m_{\pi}^{2} - s )^{2} \big \} .
\end{align}
In this interaction kernel, both the bare $\pi N \Delta$ coupling
constant $g_{\pi N \Delta}$ and $\Delta$ mass $M_{\Delta}$ are model
parameters.

As one can see, we have six model parameters altogether in the
interaction kernel $\mathcal{V}$: the low-energy constants $c_{1}$,
$c_{2}$, $c_{3}$, and $c_{4}$, the $\Delta$ bare mass $M_{\Delta}$,
and the $\pi N \Delta$ bare coupling constant $g_{\pi N \Delta}$.
They, together with the subtraction constant of the loop function, are
fixed so as to reproduce the scattering amplitude obtained in the
partial wave analysis, as explained in Sec.~\ref{sec:3A}.


\section{The loop function}
\label{app:3}

In this Appendix we summarize formulae of the loop function with the
angular momentum $L$, $G_{L} (w)$:
\begin{equation}
G_{L} ( w ) 
= i \int \frac{d^{4} q}{( 2 \pi )^{4}}
\frac{| \bm{q} |^{2 L}}{( q^{2} - m^{2} ) [ (P - q)^{2} - M^{2} ]} ,
\label{eqA:Gprop}
\end{equation}
where $P^{\mu} = ( w, \, \bm{0})$ and $m$ and $M$ are the meson and
baryon masses, respectively.  By calculating the integrals with
respect to $q^{0}$ and the solid angle of $\bm{q}$ and changing the
integral variable from $| \bm{q} |$ to $s^{\prime} = \left [
  \sqrt{\bm{q}^{2} + m^{2}} + \sqrt{\bm{q}^{2} + M^{2}} \right ]
^{2}$, we can rewrite the loop function as
\begin{equation}
G_{L} ( w ) = - \int _{s_{\rm th}}^{\infty} \frac{d s^{\prime}}{2 \pi}
\rho ( s^{\prime} ) \frac{ q(s^{\prime})^{2 L} }{s^{\prime} - s} ,
\label{eqA:GL}
\end{equation}
where $s \equiv w^{2}$, $s_{\rm th}$ is the threshold of the system,
$s_{\rm th} \equiv (m + M)^{2}$, and
\begin{equation}
\rho (s) \equiv \frac{q(s)}{4 \pi \sqrt{s}} , 
\quad 
q(s) \equiv \frac{\lambda ^{1/2} ( s , \, m^{2} , \, M^{2} )}
{2 \sqrt{s}} ,
\end{equation}
with the \Kaellen function $\lambda (x, \, y, \, z)$.  The integral in
Eq.~\eqref{eqA:GL} diverges logarithmically for $L = 0$ and worse for
$L > 0$.  However, we can ``subtract'' these divergences without
changing analytic properties of the loop function.  Namely, using an
identity
\begin{equation}
\frac{1}{s^{\prime} - s} = 
\frac{s - s_{0}}{(s^{\prime} - s) (s^{\prime} - s_{0})} 
+ \frac{1}{s^{\prime} - s_{0}} ,
\label{eqA:identity}
\end{equation}
we can reduce the order of the divergence; the first term in the
right-hand side of the identity~\eqref{eqA:identity} reduces the order
of $s^{\prime}$ of the integrand in Eq.~\eqref{eqA:GL} by one, and the
second term becomes a constant with respect to $s$ after the integral.
Iterating the substitution of this identity $L + 1$ times, we can
derive the following expression of the loop function:
\begin{align}
& G_{L} ( w ) = 
- \sum _{m = 0}^{L} \tilde{a}_{m} ( s_{0} ) ( s - s_{0} )^{m}
\notag \\
& 
- ( s - s_{0} )^{L+1} 
\int _{s_{\rm th}}^{\infty} \frac{d s^{\prime}}{2 \pi}
\rho ( s^{\prime} ) \frac{ q(s^{\prime})^{2 L} }
{( s^{\prime} - s ) ( s^{\prime} - s_{0} )^{L + 1}} .
\label{eqA:GLsubt}
\end{align}
Now this integral converges.  In the expression, $\tilde{a}_{m}$ ($m =
0$, $1$, ..., $L$) is the subtraction constant at a certain scale
$s_{0}$.  An important point is that the loop function in
Eq.~\eqref{eqA:GLsubt} has the same analytic properties, i.e., the
same dependence on $w$ as that in Eq.~\eqref{eqA:GL}.

Next we consider two special cases, $L = 0$ and $1$, which are
relevant to our study in Sec.~\ref{sec:3}.  First the $L = 0$ case
is well studied in various references.  Utilizing the dimensional
regularization for the loop function in Eq.~\eqref{eqA:Gprop} with $L
= 0$, we obtain the same structure as Eq.~\eqref{eqA:GLsubt} up to a
constant
\begin{align}
& G_{0} (w) = \frac{1}{16 \pi ^{2}} \left [ 
a ( \mu _{\rm reg} )
+ \ln \left ( \frac{M^{2}}{\mu _{\rm reg}^{2}} \right ) 
\right . 
\notag \\
& + \frac{s + m^{2} - M^{2}}{2 s} 
\ln \left ( \frac{m^{2}}{M^{2}} \right )
\notag \\ 
& - \frac{\lambda ^{1/2} (s, \, m^{2}, \, M^{2})}{s} 
\text{artanh} \left . \!
\left ( \frac{\lambda ^{1/2} (s, \, m^{2}, \, M^{2})}
{m^{2} + M^{2} - s} \right ) 
\right ] ,
\label{eqA:Gdim_explicit}
\end{align}
where $a (\mu _{\rm reg} )$ is the subtraction constant at the
regularization scale $\mu _{\rm reg}$.  Since the loop function and
scattering amplitude should be finally scale independent, any change
of the scale $\mu _{\rm reg}$ is absorbed by a change of the
subtraction constant $a( \mu _{\rm reg} )$ such that $a( \mu _{\rm
  reg}^{\prime} ) - a( \mu _{\rm reg} ) = \log ( \mu _{\rm
  reg}^{\prime \, 2} / \mu _{\rm reg}^{2} )$.  For later convenience,
here and in the following we fix the regularization scale as $\mu
_{\rm reg} = M$ and write the loop function with $\mu _{\rm reg} = M$
as $G (w ; \, a)$:
\begin{align}
& G (w ; \, a)  \equiv 
\frac{1}{16 \pi ^{2}} \left [ 
a + \frac{s + m^{2} - M^{2}}{2 s} 
\ln \left ( \frac{m^{2}}{M^{2}} \right )
\right .
\notag \\ 
& - \frac{\lambda ^{1/2} (s, \, m^{2}, \, M^{2})}{s} 
\text{artanh} \left . \!
\left ( \frac{\lambda ^{1/2} (s, \, m^{2}, \, M^{2})}
{m^{2} + M^{2} - s} \right ) 
\right ] ,
\label{eqA:Gwa}
\end{align}
where $a$ is the subtraction constant at $\mu _{\rm reg} = M$.  For
the $L = 1$ case, on the other hand, we have to take into account
another $q(s^{\prime})^{2}$ factor in the integral.  To this end we
express $q(s)^{2}$ as
\begin{equation}
q ( s )^{2} = \frac{\lambda (s , \, m^{2} , \, M^{2})}{4 s}
= \frac{s}{4} 
- \frac{m^{2} + M^{2}}{2}
+ \frac{(m^{2} - M^{2})^{2}}{4 s} .
\end{equation}
From this expression we can decompose $G_{L=1}(w)$ as
\begin{equation}
G_{1} ( w ) = \sum _{n=1}^{3} H_{n} ( w ) , 
\end{equation}
\begin{equation}
H_{1} ( w ) 
= - \frac{1}{4} 
\int _{s_{\rm th}}^{\infty} \frac{d s^{\prime}}{2 \pi}
\rho ( s^{\prime} ) \frac{ s^{\prime} }{s^{\prime} - s} ,
\end{equation}
\begin{equation}
H_{2} ( w ) 
= \frac{m^{2} + M^{2}}{2} 
\int _{s_{\rm th}}^{\infty} \frac{d s^{\prime}}{2 \pi}
\rho ( s^{\prime} ) \frac{1}{s^{\prime} - s} ,
\end{equation}
\begin{equation}
H_{3} ( w ) 
= - \frac{( m^{2} - M^{2} )^{2}}{4} 
\int _{s_{\rm th}}^{\infty} \frac{d s^{\prime}}{2 \pi}
\rho ( s^{\prime} ) \frac{1}{s^{\prime} (s^{\prime} - s)} .
\end{equation}
In order to make them converge, we need two, one, and no subtractions
for $H_{1}$, $H_{2}$, and $H_{3}$, respectively.  The first term
$H_{1}$ is calculated, with two subtraction constants $\tilde{b}_{1}$
and $a_{1}$, as
\begin{align}
H_{1} ( w ) & = - \frac{1}{4} 
\int _{s_{\rm th}}^{\infty} \frac{d s^{\prime}}{2 \pi}
\rho ( s^{\prime} ) 
\left ( 1 + \frac{s}{s^{\prime} - s} \right ) 
\notag \\ 
& = - \tilde{b}_{1} 
+ \frac{s}{4} G(w; \, a_{1}) ,
\end{align}
where a constant $\tilde{b}_{1}$ has been introduced so as to
``renormalize'' an infinite constant
\begin{equation}
\tilde{b}_{1} \equiv \frac{1}{4}
\int _{s_{\rm th}}^{\infty} \frac{d s^{\prime}}{2 \pi}
\rho ( s^{\prime} ) ,
\end{equation}
and the other integral corresponds to the $L = 0$ loop function $G(w;
\, a_{1})$, defined in Eq.~\eqref{eqA:Gwa}, with a subtraction
constant $a_{1}$.  Next, the second term $H_{2}$ is calculated in a
similar manner to the $L = 0$ loop function as
\begin{equation}
H_{2} ( w ) = - \frac{m^{2} + M^{2}}{2} G(w; \, a_{2}) ,
\end{equation}
where a subtraction constant $a_{2}$ has been introduced.  Finally,
the third term $H_{3}$ is calculated as
\begin{align}
H_{3} ( w ) 
& = - \frac{( m^{2} - M^{2} )^{2}}{4} 
\int _{s_{\rm th}}^{\infty} \frac{d s^{\prime}}{2 \pi}
\rho ( s^{\prime} ) \frac{1}{s} 
\left ( \frac{1}{s^{\prime} - s} - \frac{1}{s^{\prime}} \right )
\notag \\
& = \frac{( m^{2} - M^{2} )^{2}}{4 s} 
[ G( w ; \, 0 ) - G( 0 ; \, 0 ) ] ,
\end{align}
where we have not introduced any subtraction constant since the
integral in $H_{3}$ does not diverge.  As a consequence, we obtain the
$L = 1$ loop function $G_{1} ( w )$ as
\begin{align}
G_{1} ( w ) = & 
- \tilde{b}_{1} 
+ \frac{s}{4} G(w; \, a_{1}) 
- \frac{m^{2} + M^{2}}{2} G(w; \, a_{2}) 
\notag \\
& + \frac{( m^{2} - M^{2} )^{2}}{4 s} 
[ G( w ; \, 0 ) - G( 0 ; \, 0 ) ] .
\label{eqA:G1}
\end{align}
In this expression, we have three subtraction constants
$\tilde{b}_{1}$, $a_{1}$, and $a_{2}$, but one can easily see that
$\tilde{b}_{1}$ and $a_{2}$ are not independent.  Hence, we have two
independent subtraction constants for $G_{1} ( w )$.

In Sec.~\ref{sec:3A} we require that the $P_{11}$ unitarized partial
wave of the $\pi N$ scattering keeps the nucleon pole at the same
position as in the interaction kernel, i.e., the physical nucleon mass
$M_{N}$.  This can be achieved with the condition that the loop
function with the partial wave $L = 1$ vanishes at $w = M_{N}$.  Thus,
we require
\begin{equation}
G_{\pi N, \, L = 1} ( M_{N} ) = 0 .
\label{eqA:GpiN_L1}
\end{equation}
Then, with the condition~\eqref{eqA:GpiN_L1}, we can eliminate one of
the two independent subtraction constants in Eq.~\eqref{eqA:G1}.  As a
result, without loss of generality we can express the loop function
$G_{\pi N, \, L = 1}$ as
\begin{align}
& G_{\pi N, \, L = 1} ( w ; \, \tilde{A} ) = 
\frac{s - M_{N}^{2}}{4} \tilde{A}
+ \frac{s G_{\pi N} ( w )}{4} 
\notag \\
& 
- \frac{m_{\pi}^{2} + M_{N}^{2}}{2} G_{\pi N} ( w )
\notag \\
& + \frac{( m_{\pi}^{2} - M_{N}^{2} )^{2}}{4} 
\left [ \frac{G_{\pi N} ( w ) - G_{\pi N} ( 0 )}{s} 
+ \frac{G_{\pi N} ( 0 )}{M_{N}^{2}} \right ] ,
\end{align}
where $m_{\pi}$ is the pion mass and we have introduced the loop
function $G_{\pi N} ( w )$:
\begin{align}
& G_{\pi N} ( w ) = G ( w ; \, 0 ) - G ( M_{N} ; \, 0 ) 
\notag \\
& =
\frac{1}{16 \pi ^{2}} \left [ 
\left ( \frac{s + m_{\pi}^{2} - M_{N}^{2}}{2 s} 
- \frac{m_{\pi}^{2}}{2 M_{N}^{2}} \right )
\ln \left ( \frac{m_{\pi}^{2}}{M_{N}^{2}} \right )
\right .
\notag \\ 
& - \frac{\lambda ^{1/2} (s, \, m_{\pi}^{2}, \, M_{N}^{2})}{s} 
\text{artanh} 
\left ( \frac{\lambda ^{1/2} (s, \, m_{\pi}^{2}, \, M_{N}^{2})}
{m_{\pi}^{2} + M_{N}^{2} - s} \right ) 
\notag \\ 
& + \frac{\lambda ^{1/2} (M_{N}^{2}, \, m_{\pi}^{2}, \, M_{N}^{2})}{s} 
\text{artanh} \left . \!
\left ( \frac{\lambda ^{1/2} (M_{N}^{2}, \, m_{\pi}^{2}, \, M_{N}^{2})}
{m_{\pi}^{2}} \right ) 
\right ] .
\end{align}
which satisfies $G_{\pi N} ( M_{N} ) = 0$.  We note that $G_{\pi N, \,
  L = 1}$ contains one parameter $\tilde{A}$ as the remaining
subtraction constant.

\end{document}